\documentclass[11pt,a4paper]{article}
\pdfoutput=1 
\usepackage[utf8x]{inputenc}
\usepackage{graphicx}
\usepackage{physics}
\allowdisplaybreaks[1]
\usepackage{bm}
\usepackage{subfigure}
\usepackage[T1]{fontenc}
\usepackage{amssymb,amsmath,amsfonts}
\usepackage{stackrel}
\usepackage{float}
\usepackage{dsfont}
\usepackage{jheppub} 
\usepackage{appendix}

\newcommand{\be}{\begin{equation}}
\newcommand{\ee}{\end{equation}}
\newcommand{\bse}{\begin{subequations}}
\newcommand{\ese}{\end{subequations}}
\newcommand{\ba}{\begin{eqnarray}}
\newcommand{\ea}{\end{eqnarray}}

\newcommand{\bea}{\begin{eqnarray}}
\newcommand{\eea}{\end{eqnarray}}

\usepackage{array}
\usepackage{lineno}
\usepackage{hyperref}
\usepackage{slashbox}
\newcolumntype{L}[1]{>{\raggedright\let\newline\\\arraybackslash\hspace{0pt}}m{#1}}
\newcolumntype{C}[1]{>{\centering\let\newline\\\arraybackslash\hspace{0pt}}m{#1}}
\newcolumntype{R}[1]{>{\raggedleft\let\newline\\\arraybackslash\hspace{0pt}}m{#1}}

\begin{document}
\title{Time-dependence of the holographic spectral function: Diverse routes to thermalisation}

\author[a]{Souvik Banerjee,}
\author[b,c]{Takaaki Ishii,}
\author[d]{Lata Kh Joshi,}
\author[e]{Ayan Mukhopadhyay}
\author[d]{and P. Ramadevi}
\affiliation[a]{Van Swinderen Institute for Particle Physics and Gravity, University of Groningen, Nijenborgh 4, 9747 AG, The Netherlands}
\affiliation[b]{Department of Physics, University of Colorado, 390 UCB, Boulder, CO 80309, USA}
\affiliation[c]{Center for Theory of Quantum Matter, University of Colorado, Boulder, CO 80309, USA}
\affiliation[d]{Department of Physics, Indian Institute of Technology Bombay, Mumbai 400 076, India}
\affiliation[e]{Institut f\"{u}r Theoretische Physik, Technische Universit\"{a}t Wien, Wiedner Hauptstr.~8-10, A-1040 Vienna, Austria}

\emailAdd{souvik.banerjee@rug.nl}
\emailAdd{takaaki.ishii@colorado.edu}
\emailAdd{latamj@phy.iitb.ac.in}
\emailAdd{ayan@hep.itp.tuwien.ac.at}
\emailAdd{ramadevi@phy.iitb.ac.in}

\abstract{We develop a new method for computing the holographic retarded propagator  in generic (non-)equilibrium states using the state/geometry map. We check that our method reproduces the thermal spectral function given by the Son-Starinets prescription. The time-dependence of the spectral function of a relevant scalar operator is studied in a class of non-equilibrium states. The latter are represented by AdS-Vaidya geometries with an arbitrary parameter characterising the timescale for the dual state to transit
from an initial thermal equilibrium to another due to a homogeneous quench. For long quench duration, the spectral function indeed follows the thermal form at the instantaneous effective temperature adiabatically, although with a slight initial time delay and a bit premature thermalisation. At shorter quench durations, several new non-adiabatic features appear: (i) time-dependence of the spectral function is seen much before than that in the effective temperature (advanced time-dependence), (ii) a big transfer of spectral weight to frequencies greater than the initial temperature occurs at an intermediate time (kink formation) and (iii) new peaks  with decreasing amplitudes but in greater numbers appear even after the effective temperature has stabilised  (persistent oscillations). We find four broad routes to thermalisation for lower values of spatial momenta. At higher values of spatial momenta, kink formations and persistent oscillations are suppressed, and thermalisation time decreases. The general thermalisation pattern is globally top-down, but a closer look reveals complexities.}

\maketitle
\section{Introduction}

Holographic strongly interacting large $N$ field theories present us a great opportunity to learn about non-equilibrium dynamics of quantum many-body systems. This is because the time-evolution of multi-point Schwinger-Keldysh correlation functions, which form the quantum Martin-Schwinger hierarchy, can be expected to be solvable exactly via classical gravitational dynamics in one higher dimension involving a few fields coupled to gravity. It is indeed hard to find examples of such potentially exactly solvable interacting non-equilibrium quantum systems, unless we impose many additional symmetries or complete integrability \cite{Calabrese:2006rx,Doyon:2012bg}. 

At weak coupling and large $N$, the study of time dependence and thermalisation of two-point correlation functions has been done in a few special examples, where thermalisation has been successfully demonstrated \cite{Berges:2002wr} using the two-particle-irreducible (2PI) effective action technique (see \cite{Berges:2004yj} for a review). Nevertheless, such methods are restricted to special initial conditions and generic couplings to the environment also cannot be easily introduced. Understanding of such generic non-equilibrium evolutions is indeed desirable for many reasons. One such reason for instance, is to obtain the quantum generalisation of the theory of fluctuations in non-equilibrium classical statistical systems (see \cite{Touchette20091, 0034-4885-75-12-126001} for reviews). Another reason is to understand if there could be a sufficiently large measure of initial conditions in special quantum many-body systems which could resist thermalisation and decoherence. The power of holography suggests that we may be able to get some hints for answers to such questions, because both generic initial conditions and couplings to environment can be realised using present numerical gravity techniques (see \cite{Chesler:2013lia} for a review).

By and large, the studies on non-equilibrium aspects of the holographic correspondence, have so far been focused on the time-evolution of expectation values of gauge-invariant operators, i.e. on one-point functions. This has already led to many interesting results, for instance an explicit demonstration of emergence of hydrodynamic behaviour (see \cite{Rangamani:2009xk} for a review) -- which in turn has been instrumental in understanding qualitatively some tantalizing aspects of quantum critical systems and the quark-gluon plasma formed by heavy-ion collisions. Relatively less attention has been devoted to explicit calculations of time-dependence of multi-point correlation functions for generic initial conditions and couplings to environment. 

We will focus here on obtaining a method for finding the time-dependence of the holographic spectral function explicitly in generic non-equilibrium states. In our concrete examples, we will find radically different qualitative behaviours of the time dependence of the spectral function which reveal characteristic features of the dynamics of the non-equilibrium state, even when the one-point function of the corresponding operator does not show any distinctive transition in its behaviour as we vary external parameters. For a large class of states with homogeneous time-dependent dynamics, we will show that the holographic spectral function can follow four different patterns of thermalisation at low (spatial) momentum. We will also examine the patterns for higher momentum.

Previously, a proposal has been given in \cite{Skenderis:2008dg} for obtaining the Schwinger-Keldysh holographic partition function, by extending the Schwinger-Keldysh contour in the bulk spacetime. This proposal can in principle lead us to obtain the Schwinger-Keldysh correlation functions in non-equilibrium states of a $d-$dimensional holographic quantum field theory that can be explicitly constructed by performing a $d-$dimensional Euclidean path integral with operator insertions. This path integral leads to an explicit $(d+1)-$Euclidean dual bulk spacetime via the holographic dictionary, which should then be glued to a $(d+1)-$Lorentzian spacetime that allows us to obtain the dynamics in the real part of the Schwinger-Keldysh contour in the dual field theory.\footnote{For the thermal state, a very different realization for the holographic Schwinger-Keldysh contour has been given earlier in \cite{Maldacena:2001kr,Herzog:2002pc,Marolf:2004fy} using the causal geometry of the eternal anti-de Sitter black hole.}  At the level of two-point Schwinger-Keldysh correlation functions, it has been argued in \cite{Keranen:2014lna} that this prescription is equivalent to first obtaining the bulk Schwinger-Keldysh correlation functions in the dual (non-equilibrium) spacetime, and then extrapolating the latter to the boundary as suggested earlier in \cite{Banks:1998dd, Giddings:1999qu}. Similar arguments have also been made in \cite{Harlow:2011ke}. The method of extrapolating non-equilibrum bulk correlation functions to the boundary has been applied in several other earlier and later works \cite{CaronHuot:2011dr,Chesler:2011ds,Chesler:2012zk,Balasubramanian:2012tu,David:2015xqa,Keranen:2015mqc} for special non-equilibrium states. The major focus has been to understand the time-dependence of holographic correlations functions in a simple non-equilibrium state that undergoes instantaneous thermalisation, and which corresponds to a bulk dual represented by an AdS-Vaidya geometry that is formed by an ultra-thin homogeneous null shell of infalling matter (for applications to the quark-gluon plasma formed by heavy ion collisions see \cite{Baier:2012tc,Baier:2012ax,Muller:2012rh,Steineder:2013ana}). It can be argued that the non-equilibrium retarded correlation function for a probe operator with a large scaling dimension 
can be found using the geodesic approximation in the bulk (see \cite{Keranen:2014lna} for some not so obvious subtleties). Several works using this geodesic approximation \cite{Balasubramanian:2011ur, Aparicio:2011zy, Balasubramanian:2012tu, Ecker:2015kna} have provided some interesting hints for mechanisms of thermalisation of the spectral function. The general features of thermalisation of the retarded propagator as revealed by the geodesic approximation have been pointed out particularly in \cite{Balasubramanian:2011ur}.

However, the method of extrapolating bulk correlations to the boundary is inadequate from two points of view. Firstly, it is desirable to have a method for obtaining non-equilibrium Schwinger-Keldysh correlation functions in holography without using any additional input other than those which directly follow from the holographic correspondence itself. Secondly, it is also desirable that such a method can be used in practice to make actual computations in arbitrary non-equilibrium states. In general bulk spacetimes, one needs additional inputs for defining the bulk correlation functions themselves (as proposed in \cite{CaronHuot:2011dr,Chesler:2011ds} for instance), whose general applicability far away from equilibrium is unclear. On the other hand, the bulk extension of the Schwinger-Keldysh contour as proposed in \cite{Skenderis:2008dg} is not so easy to be implemented in practice, because determining the initial data for which a Lorentzian spacetime can be smoothly continued to an Euclidean spacetime is not always straightforward.

In an earlier work \cite{Banerjee:2012uq}, some of the authors have co-developed a method to obtain non-equilibrium spectral function in generic near-equilibrium states, which can be described holographically as black branes with decaying quasinormal mode fluctuations. It was shown that using the perturbative amplitude and derivative expansions developed in \cite{Iyer:2009in,Iyer:2011qc} one can systematically extract the time-dependence using an universal generalisation of the Son-Starinets prescription, which guarantees regular backreaction of the probe field dual to the operator under consideration. Indeed this method is exactly equivalent to the method to be proposed in the present work.\footnote{In \cite{Mukhopadhyay:2012hv}, an attempt was also made to obtain the non-equilibrium Wightman function in such near-equilibrium states indirectly combining field-theoretic arguments and additional assumptions.} We will find that one can indeed extend the method of \cite{Banerjee:2012uq} even far away from equilibrium to obtain the holographic retarded propagators of arbitrary operators in generic non-equilibrium states using only the fundamental state/geometry map as the guiding principle. Numerical implementation of our method will be based on \cite{Chesler:2013lia}.

The organisation of the paper is as follows. In Section \ref{method}, we will describe our method for obtaining the holographic retarded propagator in detail and how it can be implemented in practice numerically. We will show explicitly that our method agrees with the Son-Starinets prescription at thermal equilibrium. Furthermore, we discuss the numerical implementation for a specific class of geometries that capture holographically broad features of non-equilibrium states driven by a homogeneous quench of an arbitrary duration. 

In Section \ref{Results}, we will present our results. Following the numerical computation, we find four broad patterns of thermalisation of the spectral function at spatial momenta lower than the initial temperature. These patterns have very novel features which cannot be found in the geodesic approximation. As the modulus of the spatial momentum ($\vert \mathbf{k}\vert$) is increased we observe smooth changes in the thermalisation patterns. For instance,
the thermalisation time corresponding to any quench duration decreases with increase in $\vert \mathbf{k}\vert$.


In the concluding section, we discuss general lessons that can be learnt from our results. We will also formulate some broad questions, and discuss how we can address them in future works. The appendix provides some supporting results.

\section{Methodology and a simple implementation}\label{method}
\subsection{The prescription}

The retarded propagator is simply related to linear response even when the system is far away from equilibrium. Let us consider a $d$-dimensional system perturbed from an initial state $\ket{\Psi}$ by an external perturbation. We denote by $H_\mathrm{T}$ the total Hamiltonian given by
\begin{equation}
H_{\rm T} = H + \Delta H,
\end{equation}
where $H$ is the unperturbed Hamiltonian and $\Delta H$ is a perturbation that involves coupling of an external source $f(\mathbf{x},t)$ to an operator $O(\mathbf{x},t)$:
\begin{equation}
\Delta H(t) =\gamma \int {\rm d}^{d-1}x \, O(\mathbf{x},t) f(\mathbf{x},t),
\end{equation}
with $\gamma$ being a small dimensionless parameter denoting the strength of the perturbation. It is assumed that $\Delta H(t)$ can be switched on \textit{adiabatically} so that it does not affect the initial state $\ket{\Psi}$. Then it follows that the time evolution of the expectation value of the operator $O$ (in the Heisenberg representation with the original Hamiltonian $H$) is given by:
\begin{eqnarray}
\expval{O(\mathbf{x},t)}{\Psi}_\gamma - \expval{O(\mathbf{x},t)}{\Psi}_{\gamma=0} &=& \gamma\int{\rm d}^{d-1}x' \, \int_{-\infty}^t{\rm d}t'\, G_{\rm R}\left(\mathbf{x},t; \mathbf{x}',t'\right) f\left( \mathbf{x}',t'\right) \nonumber\\&&+ \mathcal {O}(\gamma^2),
\label{linear_response_relation}
\end{eqnarray}
where the retarded propagator $G_{\rm R}\left(\mathbf{x},t; \mathbf{x}',t'\right)$ is defined as:
\begin{equation}
G_{\rm R}\left(\mathbf{x},t; \mathbf{x}',t'\right) =-i \theta(t-t') \langle{\Psi}\vert\left[O\left(\mathbf{x},t\right),O\left(\mathbf{x}',t'\right)\right]\vert{\Psi}\rangle.
\end{equation}

Our method for computing the holographic retarded propagator implements the computation of causal response described above using the state/geometry map of holography, which is the tenet that corresponding to every solution without naked singularity in the theory of gravity there exists a state in the dual field theory, and vice versa. The chief novelty of our method involves
\begin{itemize}
\item implementation of adiabatic switching of the source, and
\item understanding of the right initial conditions to be imposed in the bulk when this additional source is switched on.
\end{itemize}
Regarding the first point, we show that we can take a suitable numerical limit where the source is a delta function in time, implying that not only the source but also \textit{all} its time derivatives vanish within desired numerical accuracy at the initial time. This automatically implies right initial conditions in the bulk. The response then directly leads us to the retarded propagator. Our method thus differs from previous ones which employ calculation of the bulk-to-bulk retarded propagator and extrapolating it to the boundary (as followed in \cite{Giddings:1999qu}, \cite{Keranen:2014lna}, etc.), or establishing an analogue of Schwinger-Keldysh contour in the bulk (as proposed in \cite{Skenderis:2008dg}). Nevertheless, our method which is generally applicable to any non-equilibrium state should be equivalent to any other alternative method whenever the latter is applicable. In particular, we will be able to reproduce the retarded propagator obtained via Son-Starinets  prescription \cite{Son:2002sd} in thermal equilibrium.

In what follows, we will review the state/geometry map as it is implemented in numerical holography, and state our prescription precisely with full generality. The expert may like to skip the remaining part of this discussion and directly move to the next subsection, where we implement our prescription in a specific simple example, and then come back to the present subsection as the generalisation of our method could be obvious. However, a reader who is not fully familiar with the methods of characteristics as it is applied to numerical holography can benefit from the following brief review and the description of how our prescription can be implemented in arbitrary non-equilibrium states.

In order to make a precise statement of our prescription for the holographic retarded propagator, it is useful to recall the state/geometry map of holography first and how it is implemented via method of characteristics \cite{Chesler:2008hg}. We will follow the notations of \cite{Chesler:2013lia}. For the sake of simplicity, let us assume that we are discussing a sector of states at strong coupling and large $N$ limit in which the energy-momentum tensor $t^{\mu\nu}$ and a scalar operator $O$ are the only single-trace operators that have non-vanishing expectation values. In this case, we may restrict ourselves to studying the dynamics of classical Einstein's gravity and a minimally coupled scalar field in the $(d+1)-$bulk dual.

Let $X$ denote the $(d+1)$-bulk coordinates collectively, and then we split them as a radial coordinate $r$, a time coordinate $t$ and spatial field theory coordinates $x^i$, with $i=1, \cdots, d-1$. For the sake of convenience, we choose an Eddington-Finkelstein type gauge in which we can write the bulk spacetime metric in the form: 
\begin{equation}\label{bulkmetric}
{\rm d}s^2 = -2 \frac{l^2}{r^2}\,{\rm d}t\left({\rm d}r +A(X) {\rm d}t+F_i(X){\rm d}x^i\right) + \frac{l^2}{r^2}H_{ij}(X){\rm d}x^i {\rm d}x^j\, .
\end{equation}
The action of the Einstein-scalar theory is
\begin{equation}
S = \frac{1}{2 \kappa^2} \int \mathrm{d}^{d+1}x \sqrt{-G} \left(R-2\Lambda - \frac{1}{2} G^{MN} \partial_M \Phi \partial_N \Phi + V(\Phi) \right).
\end{equation}
The scalar field $\Phi(X)$ is generically a function of all the bulk coordinates. In an asymptotically AdS spacetime, the leading asymptotic ($r \rightarrow 0$) behaviour of the bulk metric specifies the background metric $g^{(\rm b)}_{\mu\nu}$ on which the field theory is living:
\begin{equation}\label{bmetric}
\lim_{r\rightarrow 0}  2A(X) = -g^{(\rm b)}_{00}(t, x^i), \quad \lim_{r\rightarrow 0} F_i(X) = -g^{(\rm b)}_{0i}(t, x^i), \quad \lim_{r\rightarrow 0} H_{ij}(X) = g^{(\rm b)}_{ij}(t, x^i)
\end{equation}
In this paper, for the sake of simplicity we will assume this to be flat Minkowski space, and therefore:
\begin{equation}\label{metric-source}
g^{(\rm b)}_{00} = -1, \quad g^{(\rm b)}_{0i} = 0, \quad {\rm and} \quad g^{(\rm b)}_{ij} =\delta_{ij}. 
\end{equation}
Furthermore, the asymptotic behaviour of $\Phi(X)$ gives the source for the operator, $J(t, x^i)$, via:
\begin{equation}\label{scalar-source}
J(t, x^i) = \lim_{r\rightarrow 0}\, r^{\Delta -d} \Phi (X).
\end{equation}
Above $\Delta$ is the scaling dimension of the single-trace operator $O$, and this determines the mass of the dual field $\Phi$. If $J(t, x^i)$ is non-vanishing, it implies a Hamiltonian perturbation of the dual field theory by $JO$ -- this is often how a non-equilibrium state is realised experimentally. The asymptotic fall-off of $A(X)$, $F_i(X)$ and $H_{ij}(X)$ gives us the various components of $\langle t_{\mu\nu}(t, x^i)\rangle$, i.e. the expectation value of the energy-momentum tensor in the dual state, and similarly that of $\Phi(X)$ gives us $\langle O(t, x^i)\rangle$, i.e. the expectation value of the dual scalar operator. The latter can be extracted from the gravitational solution using the standard holographic dictionary.

In order to specify the dual state, we furthermore require to provide initial conditions in addition to the sources ($g^{\rm (b)}_{\mu\nu}$ and $J$) so that we can determine the gravitational solution uniquely. These are given by the following functions at the initial time hypersurface $t= t_{\rm in}$ (see \cite{Chesler:2013lia} for example):\footnote{Note that the $(d+1)-$variables $A(X)$, $F_i(X)$ and ${\rm det}(H)$ can be regarded as auxiliary variables -- their complete radial profile on the initial time hypersurface are also generated from the initial data. Also the initial radial profile for $\mathrm{d}_+\Phi  \equiv (\partial_t + A(X) \partial_r) \Phi$ is generated by the Klein Gordon equation from that of $\Phi(X)$.}
\begin{equation}\label{initial-data}
\hat{H}_{ij}(r, x^i) = \frac{H_{ij}}{{\rm det}(H)}, \quad \langle t_{00}(x^i) \rangle, \quad \langle t_{0i}(x^i)\rangle, \quad {\rm and}\quad \Phi(r, x^i).
\end{equation}
For appropriate initial conditions, the solutions of classical gravity will not have naked singularities in the bulk. Such solutions of classical gravity will correspond to states in the dual field theory. In case of most such initial conditions, the  exact analytic solutions cannot be readily obtained. Nevertheless, one can numerically obtain them \cite{Chesler:2013lia}. In order to check if a given initial condition is admissible or not, meaning that whether it does or does not lead to naked singularities, we will require to obtain the corresponding numerical solution over a suitable computational domain (i.e. for $0<r< r_{\rm c}$, with $r_{\rm c}$ being a suitable cut-off value of the radial coordinate) and then check if a regular apparent horizon lies within this computational domain (i.e. within $0<r< r_{\rm c}$). \footnote{The apparent horizon always lies behind (i.e. farther from the boundary than) the event horizon coinciding with the latter only at late time (see Figs. \ref{fig:hrzncpr5} and \ref{fig:hrzncpr001} for the case of AdS-Vaidya geometries). Therefore, finding the apparent horizon within the computational domain ensures absence of naked singularities.}

Let us now consider how we should obtain the retarded propagator of an operator in a given non-equilibrium state that corresponds to a solution in gravity with specific initial data (\ref{initial-data}). To begin with, let us consider the case when the concerned operators are $t^{\mu\nu}$ and the scalar operator $O$, i.e. those operators which have non-vanishing expectation values in the dual state. Obviously, we need to then perturb the field theory with additional weak sources:\footnote{We assume that there is no matter anomaly in the ground state (i.e. no contribution to conformal anomaly coming from non-trivial source $J(t, x^i)$ coupling to the scalar operator) so that ${\rm tr}(t) = t^\mu_{\phantom{\mu}\mu} = 0$ -- otherwise there will be an additional contribution to $\Delta H(t)$ proportional to ${\rm tr}(H)$.} 
\begin{equation}
\Delta H(t) = \gamma_1 \int {\rm d}^{d-1}x \, h_{\mu\nu}(t, x^i) t^{\mu\nu}(t, x^i) + \gamma_2 \int {\rm d}^{d-1}x \, f(t, x^i) O(t, x^i).
\end{equation}
Holographically, this implies that the leading asymptotic behaviour on the gravity side should be modified as follows:
\begin{equation}\label{add-source1}
g^{\rm (b)}_{\mu\nu} (t, x^i)= \eta_{\mu\nu} + \gamma_1 \,h_{\mu\nu}(t, x^i),
\end{equation}
and furthermore
\begin{equation}\label{add-source2}
\lim_{r\rightarrow 0} \, r^{\Delta -d}\, \Phi (X) = J(t, x^i) + \gamma_2\, f(t, x^i).
\end{equation}
Our crucial observation, which we will confirm numerically in the next subsection is that we can indeed implement \textit{adiabatic switching} of the sources above.\footnote{This adiabatic switching of the additional sources $f$ and $h_{\mu\nu}$ appearing in $\Delta H(t)$ should not be confused with the (non-)adiabaticity of the non-equilibrium state in which the retarded propagator is to be evaluated. The latter is decided by the external source $J$ in \eqref{add-source2} which is part of the Hamiltonian $H$ itself -- $J$ can be designed to drive instantaneous/slow quench dynamics as we will discuss in the next subsection with an example.} This means that \textit{at initial time $t= t_{\rm in}$, the additional sources $h_{\mu\nu}(t, x^i)$ and $f(t, x^i)$ and their time derivatives to all orders can be adjusted to be arbitrarily small with desired numerical accuracy. As a result, the initial data (\ref{initial-data}) can be kept unchanged so that they will lead to gravitational solutions without naked singularities, even in the presence of the additional sources.} 

Note that, it is not sufficient to make $h_{\mu\nu}(t, x^i)$ and $f(t, x^i)$ vanish at $t=t_{\rm in}$, we also need to ensure their time-derivatives to all orders vanish too with desired accuracy.  Recall that the Fefferman-Graham expansion makes the higher order time derivatives of sources appear in the radial profile of the bulk fields on-shell (on the equal-time hypersurface). Therefore, it can be expected that only if the additional sources and their time derivatives vanish at initial time, we can be sure about the original initial conditions to still remain admissible even though sources do get turned on in the future. The latter simply follows from causality, because the additional sources, which are adiabatically switched on in the future, cannot affect the initial conditions. 

We will show in the following subsection that the above adiabatic turning on of sources for switching on the Hamiltonian perturbation $\Delta H(t)$ can be achieved in holography using a sufficiently narrow Gaussian time-profile (for $h_{\mu\nu}$ and $f$ in the specific case under current discussion). Furthermore, we will also show that when the sufficiently narrow Gaussian time-profile is appropriately normalised, it mimics a delta function source to any desired degree of accuracy for computing the response at any arbitrary time. This will simplify the numerical computation of the retarded propagator. 

In order to extract the retarded propagator, we simply compute (numerically) the new gravitational asymptotically AdS solution with \textit{same} initial conditions \eqref{initial-data}, and sources (\ref{add-source1}) and \eqref{add-source2} where the additional pieces $h_{\mu\nu}$ and $f$ are adiabatically switched on as mentioned above. From the new solution, we can extract the new expectation values $\langle t_{\mu\nu}(t, x^i)\rangle_{\gamma_1,\gamma_2}$ and $\langle O(t, x^i)\rangle_{\gamma_1,\gamma_2}$ using the standard holographic dictionary. Then it follows that:
\begin{eqnarray}\label{extraction}
\langle O(t, x^i)\rangle_{\gamma_1,\gamma_2} -\langle O(t, x^i)\rangle &=& \gamma_1\int{\rm d}^{d-1}x'  \int_{-\infty}^t{\rm d}t'\,G_{\rm R}^{OT\, \mu\nu}\left(\mathbf{x},t; \mathbf{x}',t'\right) h_{\mu\nu}\left( \mathbf{x}',t'\right) \nonumber\\&&
+\gamma_2\int{\rm d}^{d-1}x'  \int_{-\infty}^t{\rm d}t'\,G_{\rm R}^{OO}\left(\mathbf{x},t; \mathbf{x}',t'\right) f\left( \mathbf{x}',t'\right)\nonumber\\&&+ \mathcal{O}\left(\gamma_1^2, \gamma_2^2, \gamma_1\gamma_2\right) , \nonumber\\
\langle t^{\mu\nu}(t, x^i)\rangle_{\gamma_1,\gamma_2} -\langle t^{\mu\nu}(t, x^i)\rangle &=& \gamma_1\int{\rm d}^{d-1}x'  \int_{-\infty}^t{\rm d}t'\,G_{\rm R}^{TT\, \mu\nu\rho\sigma}\left(\mathbf{x},t; \mathbf{x}',t'\right) h_{\rho\sigma}\left( \mathbf{x}',t'\right) \nonumber\\&&
+\gamma_2\int{\rm d}^{d-1}x'  \int_{-\infty}^t{\rm d}t'\,G_{\rm R}^{TO\, \mu\nu}\left(\mathbf{x},t; \mathbf{x}',t'\right) f\left( \mathbf{x}',t'\right)\nonumber\\&&+ \mathcal{O}\left(\gamma_1^2, \gamma_2^2, \gamma_1\gamma_2\right),
\end{eqnarray}
where $G_R$ are the following retarded propagators in the dual field theory:
\begin{eqnarray}
G_{\rm R}^{OO}\left(\mathbf{x},t; \mathbf{x}',t'\right) &=& -i \theta(t-t') \langle\left[O\left(\mathbf{x},t\right),O\left(\mathbf{x}',t'\right)\right]\rangle, \nonumber\\
G_{\rm R}^{TO\, \mu\nu}\left(\mathbf{x},t; \mathbf{x}',t'\right) &=&  -i \theta(t-t') \langle\left[t^{\mu\nu}\left(\mathbf{x},t\right),O\left(\mathbf{x}',t'\right)\right]\rangle,\nonumber\\
G_{\rm R}^{OT\, \mu\nu}\left(\mathbf{x},t; \mathbf{x}',t'\right) &=&  -i \theta(t-t') \langle\left[O\left(\mathbf{x},t\right),t^{\mu\nu}\left(\mathbf{x}',t'\right)\right]\rangle,\nonumber\\
G_{\rm R}^{TT\, \mu\nu\rho\sigma}\left(\mathbf{x},t; \mathbf{x}',t'\right) &=&  -i \theta(t-t') \langle\left[t^{\mu\nu}\left(\mathbf{x},t\right),t^{\rho\sigma}\left(\mathbf{x}',t'\right)\right]\rangle.
\end{eqnarray}
Since we know $\langle t_{\mu\nu}\rangle_{\gamma_1,\gamma_2}$ and $\langle O\rangle_{\gamma_1,\gamma_2}$, and have already specified $h_{\mu\nu}$ and $f$, we can readily extract the dual retarded propagators in the small $\gamma_1$ and $\gamma_2$ limit. The explicit numerical procedure will be discussed in the following subsection. It is quite clear that going to higher orders in $\gamma_1$ and $\gamma_2$, we can extract fully causal 3-point and other multi-point correlation functions.

Equivalently, we can also solve for the \textit{linearised} fluctuations of the bulk fields $\delta G_{MN}$ and $\delta \Phi$ with initial conditions:
\begin{equation}\label{initial-data2}
\delta\hat{H}_{ij}(r, x^i) = \delta \langle t_{00}(x^i) \rangle = \delta \langle t_{0i}(x^i)\rangle = \delta \Phi(r, x^i) = 0
\end{equation}
at $t = t_{\rm in}$, and boundary sources:
\begin{eqnarray}
\lim_{r\rightarrow 0} \, r^{\Delta -d}\, \delta \Phi (X) &=&  \gamma_2\, f(t, x^i),\nonumber\\
\delta g^{\rm (b)}_{\mu\nu}(t, x^i) &=& \gamma_1  h_{\mu\nu}(t, x^i)
\end{eqnarray}
about the background gravitational solution given by initial conditions (\ref{initial-data}) in the presence of the sources (\ref{metric-source}) and (\ref{scalar-source}). In that case, we can readily extract the dual retarded propagators using \eqref{extraction} by setting $\gamma_1 = \gamma_2 =1$ (which will be admissible because we are restricted now to the linear limit). The detailed numerical method will be presented in the following subsection. The above linearisation is certainly good enough for extracting the causal 2-point correlation function, i.e. the retarded propagator.

Clearly we can apply the same procedure for obtaining the retarded propagator of an operator $\tilde{O}$ that has vanishing expectation value in the non-equilibrium state dual to the gravitational solution obtained from the initial conditions (\ref{initial-data}) in the presence of the sources (\ref{metric-source}) and (\ref{scalar-source}). We simply study the linearised equation of motion for the dual bulk field $\tilde\Phi$ with initial condition
\begin{equation}
\tilde\Phi(r, x^i) = 0,
\end{equation}
at $t= t_{\rm in}$, in presence of a source
\begin{equation}
\lim_{r\rightarrow 0} \, r^{\tilde\Delta -d}\, \tilde\Phi (X) =  \tilde\gamma \tilde f(t, x^i)
\end{equation}
that is switched on adiabatically. From the solution we extract $\langle\tilde{O}\rangle_{\tilde\gamma}$ and then the retarded propagator $G_R^{\tilde{O}\tilde{O}}\left(\mathbf{x},t; \mathbf{x}',t'\right)$ as discussed before. 

Although we will restrict ourselves to the retarded propagator, it is to be noted that the fully causal multi-point correlation functions involving operators $\tilde{O}$ which have vanishing expectation values and those which do not can also be extracted from the full non-linear solution of the gravitational equations. In this case we switch on all sources adiabatically and keep all initial conditions unaltered. The latter implies that the initial conditions for the bulk fields that vanish in the original solution should be kept trivial, and those for the other bulk fields are kept the same as in the original solution dual to the state in which we are evaluating the correlation functions. We will call operators $\tilde{O}$ which have vanishing expectation values \textit{probe operators} from now on.

The other advantage of our method is that we can implement it even when the gravitational solutions are not known analytically, but numerically. In the following subsection, we will discuss a simple example involving a class of background gravitational solutions which are time-dependent and can be known analytically. The retarded correlation function of a probe scalar operator will be found to exhibit diverse qualitative behaviour that will reveal the characteristic features of dynamics of the bulk spacetime.

\subsection{A simple example}
\label{example}

Non-equilibrium states which can be easily realised in experiments are those which result from an external perturbation that is switched on and off adiabatically, resulting in the system being driven out of equilibrium and then eventually settling down to a new equilibrium. Gravitational solutions dual to such non-equilibrium states can be also obtained numerically. In this work, we will study a simple class of gravitational backgrounds which give a rough approximation to such dual non-equilibrium states \cite{Bhattacharyya:2009uu}. In the concluding section, we will point out that indeed our results here are also necessary to develop intuition about the time-dependence of the retarded propagator in the exact non-equilibrium background, to which we will turn our attention to in the future. We will focus on $2+1$-dimensional holographic quantum many-body systems, and therefore $3+1$-dimensional bulks.

The simple gravitational backgrounds on which we will focus our attention here are given by the metric\footnote{Note that in contrast to \eqref{bulkmetric}, the function $A$ below is $-G_{tt}$ and not $-G_{tt}/2$. We define $A$ below this way so that it coincides with the standard black hole blackening function.}
\begin{equation}\label{AdS-Vaidya-gen}
{\rm d}s^2 = \frac{l^2}{r^2} \left(-2\,{\rm d}t\,{\rm d}r - A(r,t)\, {\rm d}t^2 + {\rm d}x^2 + {\rm d}y^2\right),
\end{equation}
where $A(r,t)$ has the form of the standard black brane blackening function
\begin{equation}
A(r,t) = 1 - M(t) \,\frac{r^3}{l^2},
\end{equation}
with a time-dependent mass function
\begin{equation}\label{M-t}
M(t) = M_{\rm in} + \frac{\Delta M}{2}\left( 1 + \tanh \frac{t}{\alpha}\right).
\end{equation}
These gravitational backgrounds are known in the literature as AdS-Vaidya spacetimes. The mass of the black hole in these spacetimes starts with the value $M_{\rm in}$ in the far past $t=-\infty$, and then increases monotonically to $M_{\rm f}$ in the far future $t=\infty$, with $M_{\rm f} = M_{\rm in} +\Delta M$. Without loss of generality, we have arranged that the mass function attains the central value $(M_{\rm in} + M_{\rm f})/2$ at $t=0$. The parameter $\alpha$ controls the time scale over which the mass function changes significantly. Note,
\begin{equation}
\lim_{\alpha \rightarrow 0} M(t) = M_{\rm in} + \Delta M \, \theta(t).
\end{equation}
This corresponds to the case when an ultra-thin homogeneous null shell infalls into a black hole changing its mass from $M_{\rm in}$ to $M_{\rm f}$. From the dual field theory point of view, thus the limit $\alpha \rightarrow 0$ corresponds to a very rapid quench. The time-dependence of the retarded propagator in such backgrounds has been studied in the literature mostly for the case $M_{\rm in} = 0$ (see the discussion in the introduction). When $\alpha$ is finite, the gravitational background  \eqref{AdS-Vaidya-gen} is then generated by an infalling homogeneous null shell of effective width $2\alpha$. The limit $\alpha\rightarrow\infty$ corresponds to an adiabatic process of transition of the black hole mass. We call $\alpha$ the \textit{quenching parameter}.

\begin{figure}[t]\label{app-event-horizon}
\centering
\subfigure[$\alpha=5$]{\includegraphics[height=3.2cm, keepaspectratio]{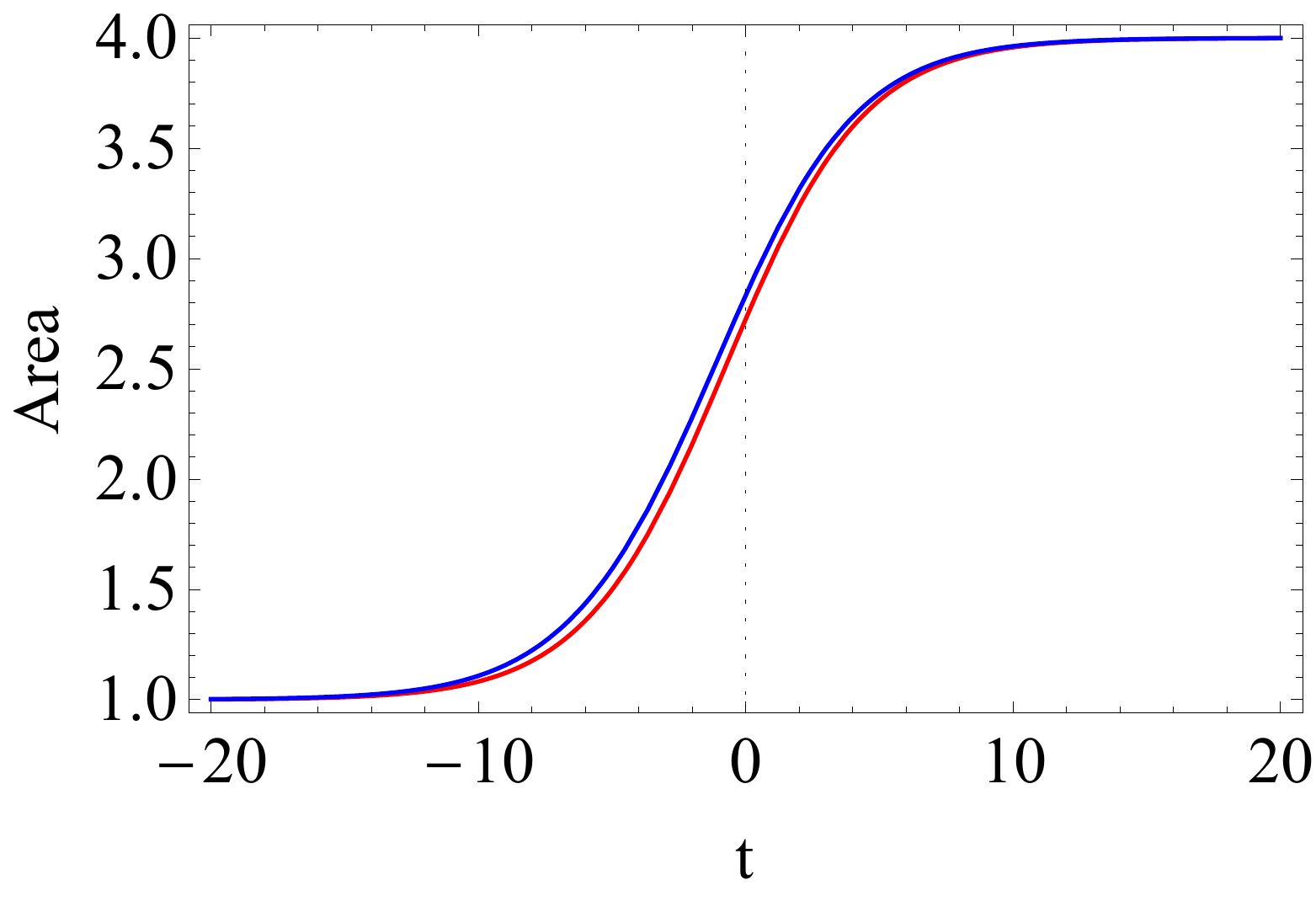}\label{fig:hrzncpr5}}
\subfigure[$\alpha=0.01$]{\includegraphics[height=3.2cm]{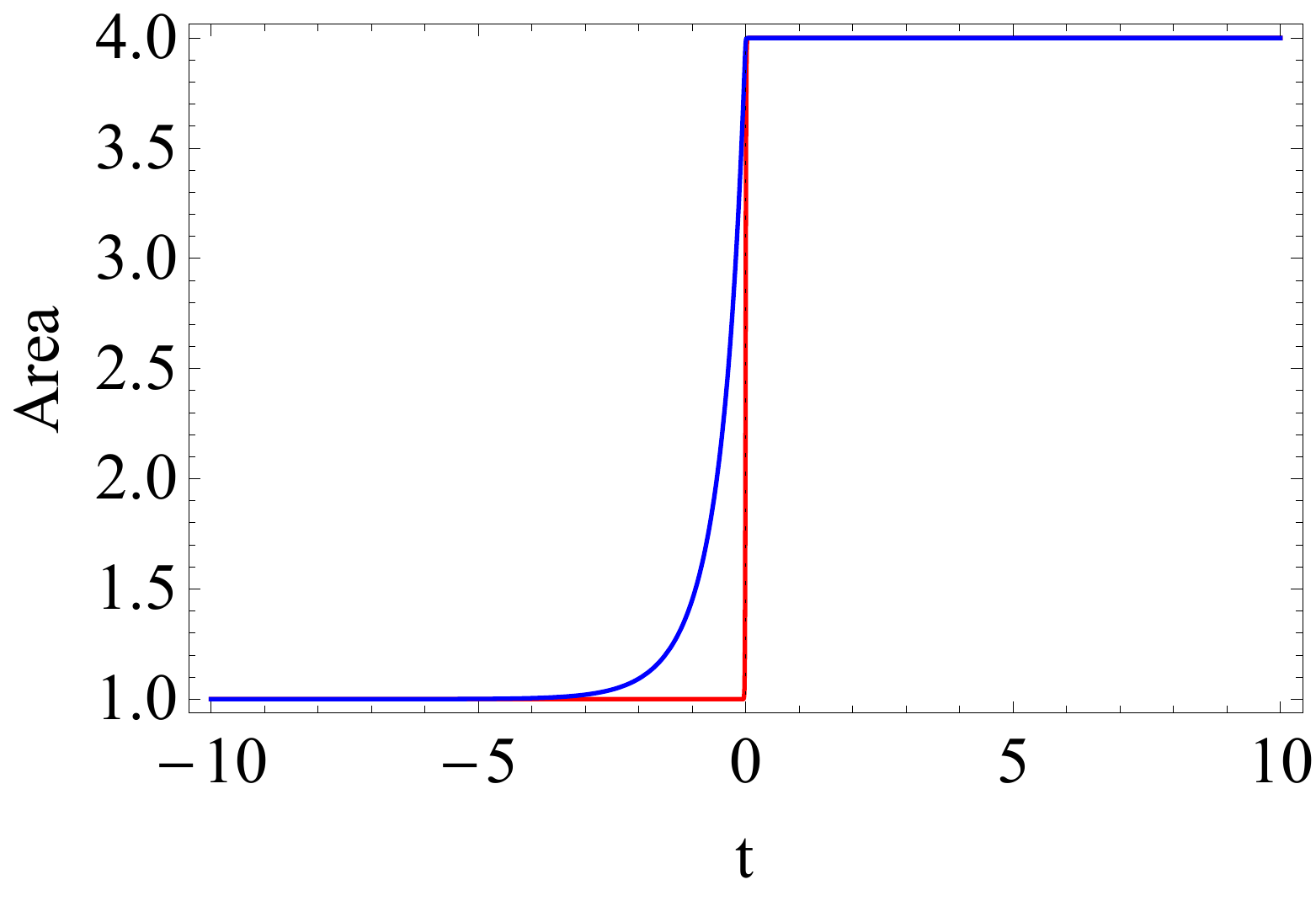}\label{fig:hrzncpr001}}
\subfigure[Event horizons]{\includegraphics[height=3.25cm]{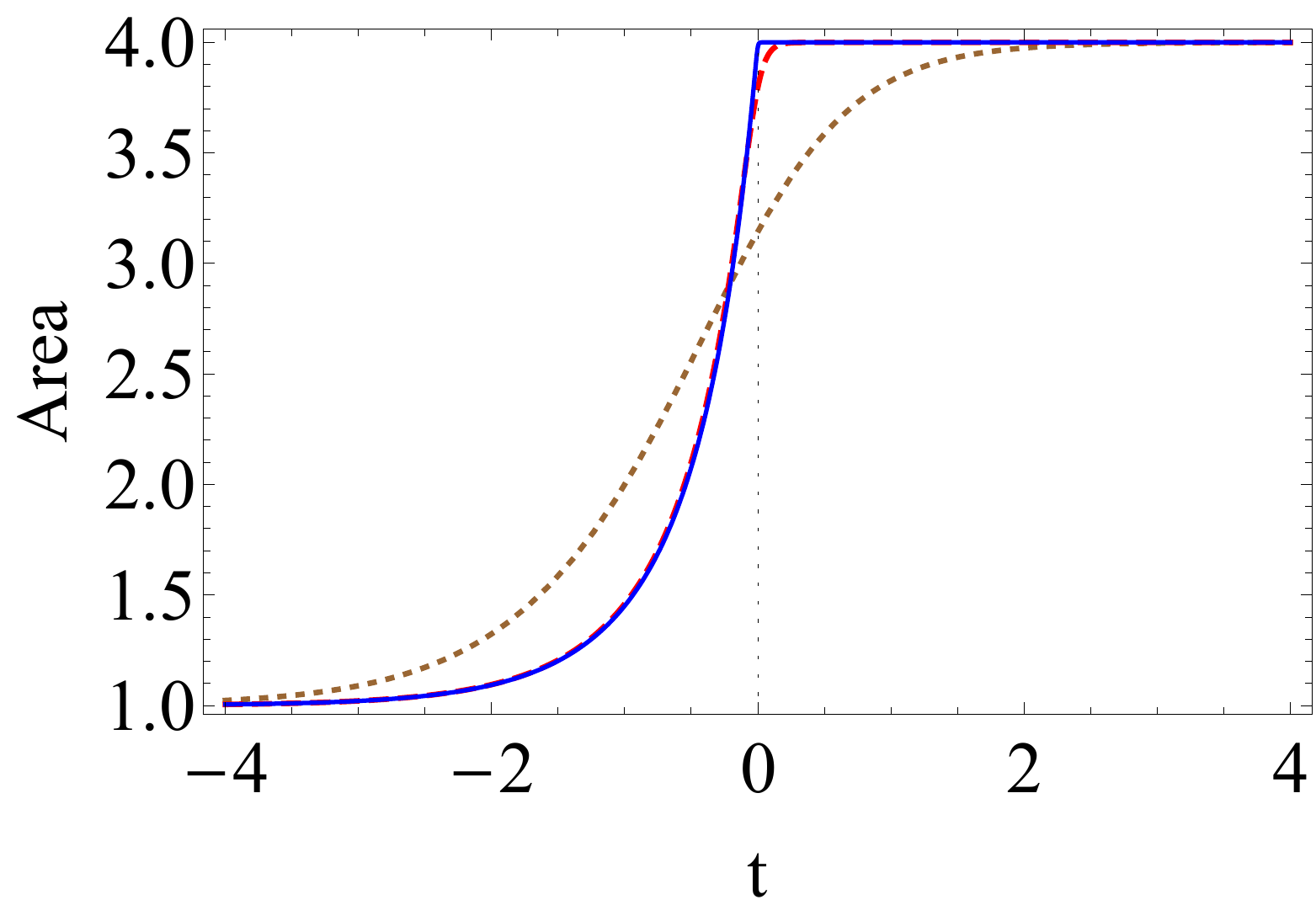}\label{fig:hrzncprcpr}}
\caption{Horizons for different $\alpha$. In Figs.~\ref{fig:hrzncpr5} and \ref{fig:hrzncpr001}, apparent (red) and event (blue) horizons are plotted for $\alpha=5$ and $\alpha=0.01$, respectively. In figure~\ref{fig:hrzncprcpr}, event horizons are compared for $\alpha =$1 (dotted brown), 0.1 (dashed red), and 0.01 (solid blue). The last two cases almost coincide.}
\label{fig:hrzncpr}
\end{figure}

We recall that in the static background where $M(t)=M$, the horizon of the black brane is located at $r_{\rm H} = l^{2/3} M^{-1/3}$. The event and apparent horizons typically do not coincide in non-stationary gravitational backgrounds. In the AdS-Vaidya spacetimes \eqref{AdS-Vaidya-gen}, the location of the apparent horizon $r_\mathrm{AH}(t)$ is given by the solution of $A(r_\mathrm{AH}(t),t)=0$. The location of the event horizon $r_\mathrm{EH}(t)$ is obtained by solving the null geodesic equation $\dot{r}_\mathrm{EH} = - A(r_\mathrm{EH}(t),t)/2$ subject to the future boundary condition $r_\mathrm{EH}(t \to \infty) = l^{2/3}M_{\rm f}^{-1/3}$, which ensures that in the far future it reproduces the event horizon of the final black hole. In Fig.\ref{fig:hrzncpr} we show the evolution of the apparent and the event horizon for various values of $\alpha$. In these plots, and for the remaining paper we set units $l=1$ and choose $M_{\rm in} = 1$, $M_{\rm f} =8$ (i.e. $\Delta M = 7$). The final temperature $T_{\rm f}$ is then twice higher than the initial temperature $T_{\rm in}$. Note that both in the far past and far future, the event and apparent horizons of \eqref{AdS-Vaidya-gen} coincide. 

The Hawking temperature of the static black brane is given by
\begin{align}
T_{\rm H} = \frac{3}{4 \pi r_{\rm H}} = \frac{3 M^{1/3}}{4 \pi l^{2/3}}.
\label{hawking_temperature}
\end{align}
We may thus think about $(3 M(t)^{1/3})/(4 \pi l^{2/3})$ as an effective instantaneous temperature $T(t)$ of the AdS-Vaidya spacetime. However, such a description is strictly valid only in the adiabatic $\alpha\rightarrow\infty$ limit.

We are going to study the time-dependence of the retarded propagator of a \textit{probe} scalar operator $O$ in the non-equilibrium state dual to the AdS-Vaidya gravitational background (\ref{AdS-Vaidya-gen}) using the prescription described in the previous subsection. This operator $O$ does not have a spontaneous expectation value in the dual state, and is dual to a bulk scalar field $\Phi$ that trivially vanishes in the background (\ref{AdS-Vaidya-gen}) before turning on the boundary source perturbation. We will also consider the case when this operator is \textit{relevant} with scaling dimension $\Delta = 2$. The latter translates to the mass of the dual bulk scalar field being given by $m^2 l^2  = -2$.\footnote{By the general AdS/CFT dictionary, the scaling dimension $\Delta$ of a scalar operator in the $d-$dimensional CFT and the mass $m$ of the dual bulk scalar field in $AdS_{d+1}$ are related by $\Delta = (d/2) + \sqrt{(d/2)^2 + m^2 l^2}$. Note in AdS there is no tachyonic instability for $m^2 l^2 \geq - (d/2)^2$. This inequality gives the Breitenlohner-Freedman bound \cite{Breitenlohner:1982bm} on the masses of the scalar fields in AdS.} 

We also assume that the bulk scalar field $\Phi$ is minimally coupled to gravity and does not couple to the matter composing the null shell that generates the AdS-Vaidya background \eqref{AdS-Vaidya-gen}. The bulk action for the scalar field is then
\begin{equation}
S = - \frac{1}{2}  \int {\rm d}^4 x \sqrt{-G} \left( G^{MN} \partial_M \Phi \partial_N \Phi - \frac{2}{l^2} \Phi^2 \right).
\label{action_scalar}
\end{equation}
Since the background \eqref{AdS-Vaidya-gen} preserves spatial homogeneity, the solution for the Klein-Gordon equation for $\Phi$ in this background can be written in the form 
\begin{equation}
\Phi(r, t, \mathbf{x}) = \Phi_0(r,t, \mathbf{k}) e^{i\mathbf{k}\cdot \mathbf{x}}.
\end{equation}
The Klein-Gordon equation then reduces to
\begin{equation}
(\mathrm{d}_{+} \Phi_0)' - \frac{\mathrm{d}_{+} \Phi_0}{r} + \frac{A}{2 r} \Phi_0' + \left(- \frac{1}{r^2} + \frac{\mathbf{k}^2}{2} \right) \Phi_0 = 0,
\label{eom_probephi}
\end{equation}
where ${}^\prime \equiv \partial_r$, and 
\begin{equation}\label{d+Phi}
\mathrm{d}_{+} \equiv \partial_t - (A/2) \partial_r 
\end{equation}
is the derivative along the outgoing null geodesic. 

Any solution of \eqref{eom_probephi} can be written asymptotically in the form
\begin{equation}
\Phi_0(r,t, \mathbf{k}) = f_0(t, \mathbf{k}) r + f_1(t, \mathbf{k}) r^2 + \mathcal{O}(r^3).
\label{phi_bdry_expansion}
\end{equation}
Applying $\mathrm{d}_{+}$ to the  equation above, we readily obtain the asymptotic form of $\mathrm{d}_{+} \Phi_0$ as below:
\begin{equation}\label{eom-EF}
\mathrm{d}_{+} \Phi_0(r,t, \mathbf{k}) = -\frac{f_0(t, \mathbf{k})}{2} + (- f_1(t, \mathbf{k})+ \dot{f}_0(t, \mathbf{k})) r + \mathcal{O}(r^2),
\end{equation}
where $\dot{} \equiv \partial_t$. According to the holographic dictionary, the external source $f(t, \mathbf{x}) = f_0(t, \mathbf{k}) e^{i\mathbf{k}\cdot\mathbf{x}}$ leads to the perturbation of the Hamiltonian of the system by
\begin{equation}
\Delta H(t) = \int {\rm d}^2 x \, f(t, \mathbf{x}) O(t, \mathbf{x}).
\end{equation}
Furthermore, the expectation value of $\langle O(t,\mathbf{x})\rangle_f =\langle O_0(t,\mathbf{k})\rangle_f \, e^{i\mathbf{k}\cdot\mathbf{x}}$ in the presence of the source $f(t,\mathbf{x})$ can be read off from the bulk solution using\footnote{We obtain the result below by changing to Fefferman-Graham coordinates $(z, \tilde{t}, x, y)$ first using
\begin{equation}
r = z  + \cdots, \quad
t = \tilde{t} - z  + \cdots.
\end{equation}
After doing this change of coordinates, the asymptotic expansion in the Fefferman-Graham coordinates $\Phi_0(z, \tilde{t}, \mathbf{k})$ is written in terms of the coefficients in the Eddington-Finkelstein coordinates \eqref{phi_bdry_expansion} as
\begin{equation}
\Phi_0(z, \tilde{t}, \mathbf{k}) = f_0(t, \mathbf{k}) z + \left(f_1(t,\mathbf{k})-\dot{f}_0(t,\mathbf{k})\right) z^2 + \mathcal{O}(z^3).
\end{equation}
The one-point function $\langle O \rangle_f$ is minus one times the coefficient of $z^\Delta$ of the Fefferman-Graham expansion times $(2\Delta -d)$ when $\log r$ terms are absent (see \cite{Skenderis:2002wp} for instance). Since in our case $\Delta = 2$ and $d=3$, we obtain \eqref{O}.
}
\begin{equation}\label{O}
\langle O_0(t,\mathbf{k})\rangle_f = - f_1(t,\mathbf{k})+\dot{f}_0(t,\mathbf{k}).
\end{equation} 

In practice, the Klein-Gordon equation \eqref{eom_probephi} can be solved using the method of characteristics. At initial time $t = t_{\rm in}$, we specify $\Phi_0(r, t, \bm{k})$. We then radially integrate \eqref{eom_probephi} to generate $\mathrm{d}_+ \Phi_0(r,t,\bm{k})$ at  $t = t_{\rm in}$ using the boundary condition $\mathrm{d}_+ \Phi_0(r = 0,t_{\rm in},\bm{k}) = -f_0(\mathbf{k}, t_{\rm in})/2$ that follows from the asymptotic expansion \eqref{eom-EF}. Then we obtain $\partial_t \Phi_0$ from $\mathrm{d}_+ \Phi_0$ using $\partial_t \Phi_0 =\mathrm{d}_+ \Phi_0 + (A/2) \Phi_0$ (recall the definition \eqref{d+Phi} of $\mathrm{d}_+$), and update $\Phi_0(r, t, \bm k)$ at the next time step $t = t_{\rm in} + \Delta t$. We further continue at the next time step by performing the radial integration in \eqref{eom_probephi} to obtain $\mathrm{d}_+ \Phi_0(r,t_{\rm in} +\Delta t,\mathbf{k})$ using the boundary condition provided by the given source $f_0(\mathbf{k}, t_{\rm in} +\Delta t)$, and so on. Thus, the unique solution for $\Phi_0$ can be obtained numerically for a given radial profile at initial time and a boundary source $f_0$ specified at all times. Here, we will use the spectral method with $30$ grid points to perform the radial integration for obtaining $\mathrm{d}_+ \Phi_0(r,t,\mathbf{k})$ at each time step. In order to update in time, we will use a fourth order Adams-Bashforth time stepper with $\Delta t= .0005$.

As discussed in the previous subsection, in order to obtain the retarded propagator $G_R^{OO}(\mathbf{x},t, \mathbf{x}', t')$ in the non-equilibrium state of interest, we need to switch on the source $f(t, \mathbf{x})$ adiabatically by designing an appropriate profile for it, then obtain $\langle O(t) \rangle_f$ by solving \eqref{eom_probephi} in the AdS-Vaidya background \eqref{AdS-Vaidya-gen}. We need to impose the initial condition $\Phi(r, x^i) = 0$ in the far past $t = -\infty$, and compute the retarded propagator from the relation:
\begin{equation}
\langle O(t, \mathbf{x}) \rangle_f = \int_{-\infty}^t {\rm d}t' \int {\rm d}^2x' \, \, G_R^{OO}(\mathbf{x},t; \mathbf{x}', t')f(t', \mathbf{x}').
\end{equation}
Note that due to spatial homogeneity of the background \eqref{AdS-Vaidya-gen}, the retarded propagator can depend on $\mathbf{x} - \mathbf{x}'$, and not $\mathbf{x}$ and $\mathbf{x}'$ separately. It is therefore convenient to study the Fourier transform of $G_R^{OO}(\mathbf{x},t; \mathbf{x}', t')$ with respect to $\mathbf{x} - \mathbf{x}'$, which we denote as $G_R^{OO}(t, t'; \mathbf{k})$ (with a slight abuse of notation). Clearly,
\begin{equation}\label{relation2}
\langle O_0(t, \mathbf{k}) \rangle_f = \int_{-\infty}^t {\rm d}t' \, \, G_R^{OO}(t, t'; \mathbf{k})f_0(t', \mathbf{k}).
\end{equation}
Remarkably, we find that it is possible to implement the delta function limit of the source, so that $f_0(t', \mathbf{k}) = \delta(t'- t_0)$. Practically, we implement this delta function as a narrow ($\sigma \rightarrow 0$) limit of the Gaussian function \footnote{Boundary source perturbation by a narrow but finite width Gaussian profile has been often used to drive the bulk scalar in numerical time evolution and compute the one point function \cite{Bhaseen:2012gg,Craps:2014eba,Ishii:2015gia,Craps:2015upq}. Scalar quenches have also been studied with other boundary source profiles \cite{,Buchel:2012gw,Basu:2012gg,Buchel:2013lla,Buchel:2013gba,Das:2014lda,Buchel:2014gta}.}
\begin{align}
f_0(t', \mathbf{k}) = \frac{1}{\sqrt{2\pi}\sigma} e^{-\frac{(t'-t_0)^2}{2\sigma^2}},
\label{eq_deltagaussfn}
\end{align}
which is  centred at $t'=t_0$ and is normalised such that
\begin{equation}
\int_{-\infty}^{\infty} {\rm d}t' f_0(t', \mathbf{k}) = 1.
\end{equation}
We will soon show that numerically it is indeed possible to take the limit $\sigma \rightarrow 0$ smoothly in \eqref{eq_deltagaussfn} for evaluating $\langle O_0(t, \mathbf{k}) \rangle_f$ (and we will also make error estimates when $\sigma$ is finite). 

Indeed if $f_0(t')$ is given by $\delta(t' - t_0)$, then we can readily read off $G_R^{OO}(t, t')$ from \eqref{relation2} using
\begin{equation}\label{relation-simple}
G_R^{OO}(t, t_0; \mathbf{k}) = \langle O_0(t, \mathbf{k}) \rangle_f.
\end{equation}
Although the integral over $t'$ runs from $-\infty$ to $t$, we can extend this integral over the entire real line, because $\langle O_0(t, \mathbf{k}) \rangle_f $ and hence $G_R^{OO}(t, t_0; \mathbf{k})$ vanishes for $t' > t$ by causality. Thus the above result indeed follows if $f_0$ is given by $\delta(t - t_0)$. Therefore, we can directly read off the retarded propagator $G_R^{OO}(t, t_0; \mathbf{k}) $ from $ \langle O_0(t, \mathbf{k}) \rangle_f$ (given as discussed earlier by \eqref{O}) for a fixed value of $t_0$. We need to vary $t_0$ though to obtain complete information about $G_R^{OO}(t, t_0; \mathbf{k}) $. Using \eqref{O}, we can also write \eqref{relation-simple} 
as
\begin{equation}\label{relation-simple-2}
G_R^{OO}(t, t'; \mathbf{k}) = -f_1(t, \mathbf{k}) + \dot{\delta}(t -t').
\end{equation}
where $ f_1$ is evaluated in presence of the source $f_0(t, \mathbf{k})  = \delta(t -t')$ and with initial condition $\Phi_0(r,\mathbf{k}, t_{\rm in}) $ vanishing in the far past.

There is another advantage of using the narrow Gaussian approximation for the Dirac delta function source \eqref{eq_deltagaussfn}, namely that it readily implements adiabatic switching of the source that is necessary for our prescription for obtaining the retarded propagator discussed in the previous subsection. The time $t_{\rm in}$ when the initial condition $\Phi(r, x^i) = 0$ is set can be chosen well before $t= t_0$ (i.e. $t_{\rm in} \ll t_0 - \sigma$) when the narrow Gaussian profile \eqref{eq_deltagaussfn} of the source has a significant non-trivial value -- the source and its time derivatives to all orders vanish to desired level of accuracy at $t = t_{\rm in}$. Furthermore, as we will show quantitatively below, indeed for the causal response $\langle O_0(t, \mathbf{k}) \rangle_f $ for $t> t_0$ (when $G_R^{OO}(t, t_0; \mathbf{k})$ should be non-vanishing), the convergence to this delta-function limit $\sigma \rightarrow 0$ is very rapid. Thus we can readily obtain $G_R^{OO}(t, t_0; \mathbf{k})$ using \eqref{relation-simple-2} for any given $t_0$ by appropriately solving the bulk Klein-Gordon equation for the bulk scalar field $\Phi$ with initial condition $\Phi(r, t_{\rm in}, x^i) = 0$ at an appropriate value of  the initial time $t_{\rm in}$.

To obtain better physical intuition about $G_R^{OO}(t, t'; \mathbf{k})$ and also to make connections with experiments, it is instructive to study the Wigner transform of $G_R^{OO}(t, t'; \mathbf{k})$ which is defined by below:
\begin{equation}
G_R^{OO}(\omega, t_{\rm av}, \mathbf{k}) = \int_{-\infty}^{\infty} {\rm d}t_{\rm rel}\, e^{i\omega t_{\rm rel}} G_R^{OO}\left(t =  t_{\rm av} + \frac{t_{\rm rel}}{2}, t' =  t_{\rm av} - \frac{t_{\rm rel}}{2} ; \mathbf{k}\right).
\end{equation}
The Wigner transform is thus a Fourier transform with respect to the relative coordinate $t_{\rm rel} = t - t'$ and then after this transformation $G_R^{OO}$ depends on the frequency $\omega$ and the average time $t_{\rm av} = (t+t')/2$. Again for the sake of notational simplicity, we denote the Wigner transformed $G_R^{OO}$ as $G_R^{OO}(\omega, t_{\rm av}, \mathbf{k})$. When $t_{\rm av} \rightarrow \mp \infty$,  $G_R^{OO}(\omega, t_{\rm av}, \mathbf{k})$ does not depend on $t_{\rm av}$, and coincides with the thermal retarded propagator $G_R^{OO}(\omega, \mathbf{k})$ with temperatures $(3 M_{\rm in, f}^{1/3})/(4 \pi l^{2/3})$ corresponding to the black brane backgrounds with masses $M_{\rm in}$ and $M_{\rm f}$ respectively, as discussed before.

We will focus mostly on the spectral function $\rho^{OO} (\omega, t_{\rm av}, \mathbf{k})$ which is experimentally measurable for example in solids using angle resolved photo electron spectroscopy (ARPES) and is defined as
\begin{equation}\label{spec-function}
\rho^{OO} (\omega, t_{\rm av}, \mathbf{k}) = -2 \, {\rm Im} \, G_R^{OO}(\omega, t_{\rm av}, \mathbf{k}).
\end{equation}
Causality implies that $G_R^{OO}(\omega, t_{\rm av}, \mathbf{k})$ is analytic in $\omega$ in the upper half complex plane, therefore the Kramers-Kronig relation
\begin{equation}
{\rm Re}\,G_R^{OO} (\omega, t_{\rm av}, \mathbf{k}) = -\frac{1}{2\pi} \mathcal{P} \int_{-\infty}^{\infty} {\rm d}\omega' \, \frac{\rho^{OO} (\omega', t_{\rm av}, \mathbf{k})}{\omega' -\omega}
\end{equation}
is valid, with $\mathcal{P}$ denoting the Cauchy principal value. Thus causality implies that the spectral function encodes complete information of the retarded propagator. 

Because $G_R^{OO}(t, t'; \mathbf{k})$ is real, we also note that, 
\begin{equation}
G_R^{OO}(-\omega, t_{\rm av}, \mathbf{k}) = {G_R^{OO}}^*(\omega, t_{\rm av}, \mathbf{k}) = G_A^{OO}(\omega, t_{\rm av}, \mathbf{k}),
\end{equation}
where $G_A$ is the advanced propagator. This implies:
\begin{equation}
\rho^{OO}(-\omega, t_{\rm av}, \mathbf{k}) = - \rho^{OO}(\omega, t_{\rm av}, \mathbf{k}).
\end{equation}
It is therefore sufficient to restrict ourselves to the region $\omega > 0$.\footnote{The reader can readily recall that for a free bosonic field $\chi$ with mass $m$, the \textit{vacuum} spectral function is given by:
\begin{equation}
\rho^{\chi\chi}(\omega, \mathbf{k}) = \frac{\pi}{2\epsilon_{\mathbf{k}}}\left(\delta(\omega -\epsilon_{\mathbf{k}}) - \delta(\omega +\epsilon_{\mathbf{k}}) \right),
\end{equation} with $\epsilon_{\mathbf{k}} = \sqrt{\mathbf{k}^2 + m^2}$.}

Since we are studying the case of a Hermitian operator dual to a real scalar field, $G_R^{OO}(t, t'; \mathbf{k})$ is real as well. Therefore, it follows from \eqref{relation-simple-2} that
\begin{equation}\label{relation-simple-3}
\rho^{OO} (\omega, t_{\rm av}, \mathbf{k}) = 2\omega + 2 \int_{-\infty}^{\infty} {\rm d}t_{\rm rel} \, \sin(\omega t_{\rm rel}) \, f_1\left(t_{\rm av} + \frac{t_{\rm rel}}{2}, \mathbf{k}\right),
\end{equation}
where $f_1$ is obtained from solving the Klein-Gordon equation in the background \eqref{AdS-Vaidya-gen} in the presence of the source $f_0 = \delta(t-t')$ and with initial condition $\Phi_0(r, t_{\rm in},\mathbf{k}) = 0$ in the far past.

\begin{figure}[t]
\centering
\subfigure[$f_1$ when $\sigma=0.01$]{\includegraphics[height=3.2cm]{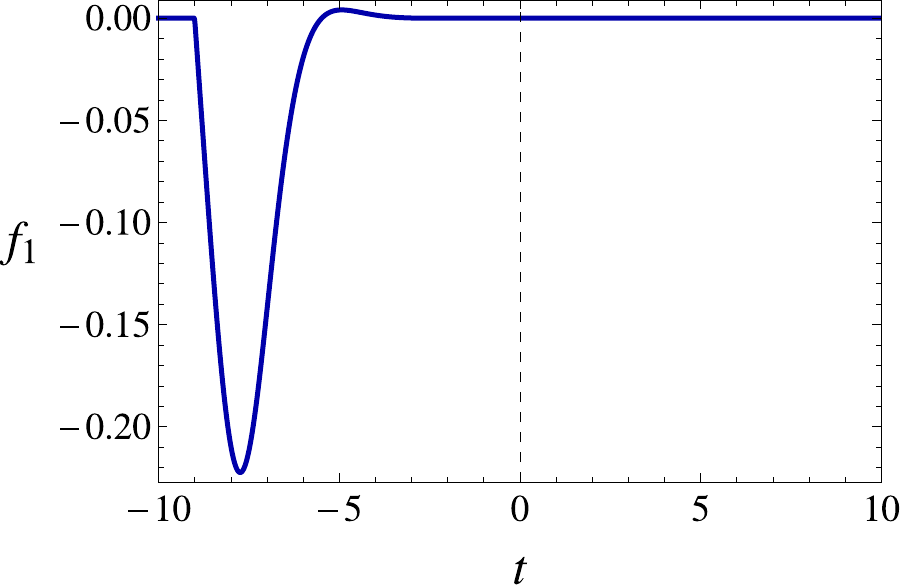}\label{fig:f1plot1}}
\subfigure[Log plot of $|f_1|$ when $\sigma=0.01$]{\includegraphics[height=3.2cm]{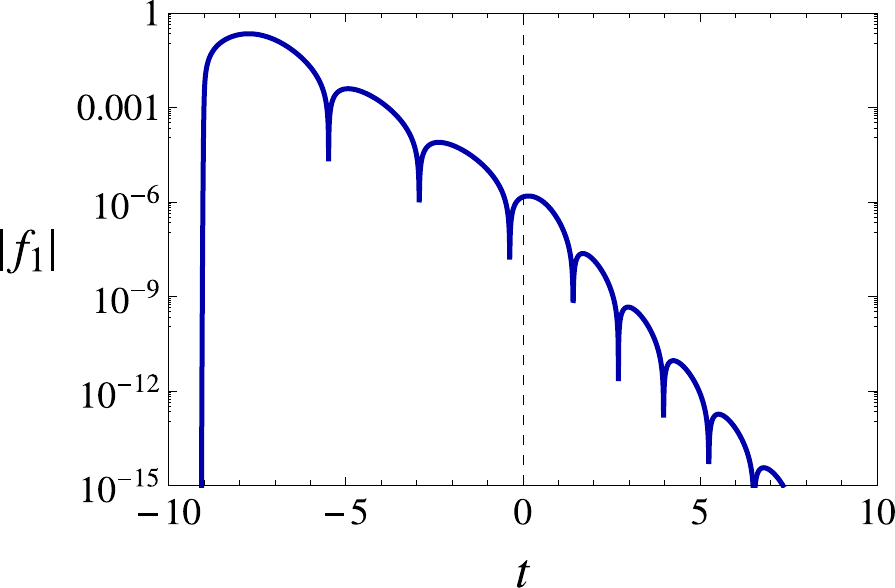}\label{fig:f1plot2}}
\subfigure[$\sigma = 0.1, \, 0.01, \, 0.001$]{\includegraphics[height=3.2cm]{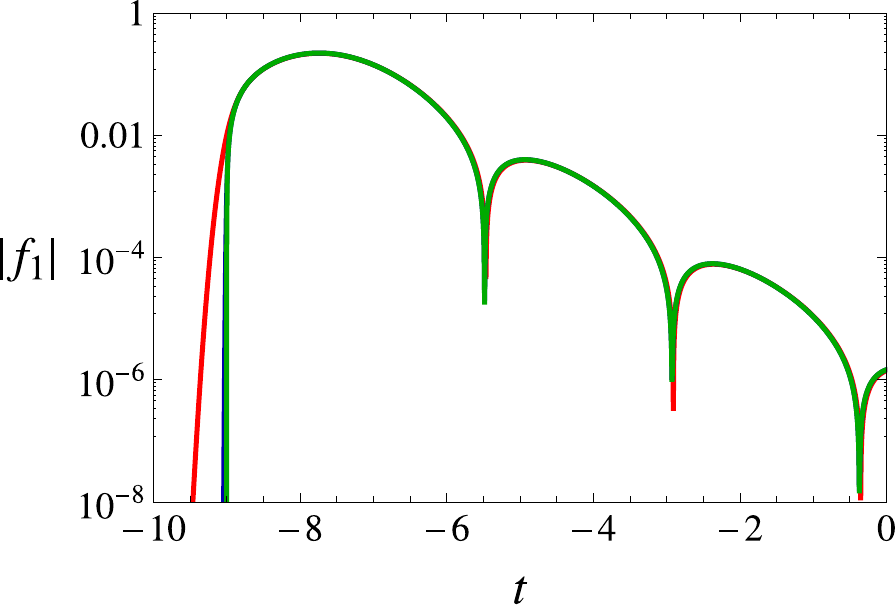}\label{fig:f1plot3}}
\caption{Results when $\alpha=0.05, \, t_0=-9, \, \mathbf{k}=0$. Red, blue and green lines correspond to $\sigma=0.1, \, 0.01, \, 0.001$ respectively in Fig. \ref{fig:f1plot3} -- note that the blue and green lines almost coincide at all times demonstrating the achievement of the delta function limit within desired numerical accuracy.}
\label{fig:f1plot}
\end{figure}

In order to get a first look at how well we can approximate the delta function source by a narrow Gaussian, we study the behaviour of $f_1(t, \mathbf{k})$. We set $\mathbf{k}=0$ and locate a narrow Gaussian perturbation at $t_0 = -9$.\footnote{Throughout this paper, we choose $M_{\rm in} = 1$ and $M_{\rm f} = 8$ for the AdS-Vaidya background geometry as mentioned before.} In Fig.~\ref{fig:f1plot1} and \ref{fig:f1plot2}, we show the result of $f_1$ for $\sigma=0.01$ when $\alpha=0.05, \, \mathbf{k}=0$. When the shell is thin, the background immediately approaches the thermal equilibrium, and hence a sharp transition from the initial to final temperatures is observed in Fig.~\ref{fig:f1plot2}. The oscillating exponential decay of the one point function away from the shell is controlled by the lowest lying quasi normal mode frequency. For $\Delta=2, \, \mathbf{k}=0$ in AdS$_4$, the value is $3 \omega/(4 \pi T_{\rm f}) = 1.228 - 1.533 i$ as can be obtained using the method of \cite{Starinets:2002br}. Our numerical results agree with this value.

By varying $\sigma$, we find that the behaviour of $f_1(t, \mathbf{k})$ converges as we decrease $\sigma$. See Fig.~\ref{fig:f1plot3} comparing $\sigma = 0.1, \, 0.01, \, 0.001$. At early time, different curves of $f_1$ do not coincide due to different widths of the corresponding sources, but at late times the ring down parts lie almost on top of each other. However, for sufficiently small $\sigma$, the curves of $f_1$ coincide well at all times. We can take the limit of delta function source by computing for several small $\sigma$ and extrapolating to $\sigma \to 0$. Using a small $\sigma$ is sufficient for practical numerical purpose.

One may easily estimate the influence of the finite width  $\sigma$ of the Gaussian profile for the source on the numerical error in finding the Wigner transform. Note that the Fourier transform of the Gaussian profile \eqref{eq_deltagaussfn} for $f_0(t, \mathbf{k})$ is:
\begin{equation}
\tilde{f}_0(\omega, \mathbf{k}) \propto e^{-\frac{\omega^2\sigma^2}{2}}.
\end{equation}
Following the above it is not hard to see that when we extract spectral function treating the source $f_0$ as a delta function in real time using the Fourier transform \eqref{relation-simple-3}, we are making significant errors only for very high values of $\omega$, i.e. for $\omega \gg \sigma^{-1}$. However, even before reaching such high values of $\omega$, we will find that $\rho(\omega, t_{\rm av}; \mathbf{k})$ asymptotes the vacuum form (with $\Delta = 2$):
\begin{equation}\label{rho-AdS}
\rho_{\rm AdS}(\omega, \mathbf{k}) = 2\, \theta \left({\rm abs}(\omega) -\vert \mathbf{k}\vert\right)\, \sqrt{\omega^2 - \mathbf{k}^2} \,{\rm sgn}(\omega),
\end{equation}
for all values of $t_{\rm av}$ when $\mathbf{k}$ is not too large. This also requires that we should take the initial and final black hole masses $M_{\rm in, f} \ll \sigma^{-1}$, because the mass parameters control the frequency above which time-dependent spectral function asymptotes to the vacuum form \eqref{rho-AdS}, at least for low values of $\vert \mathbf{k} \vert$. 

Note in a conformal field theory there is no intrinsic scale, so only ratios of various mass parameters, time scales, etc. affect physical results. Therefore, without loss of generality we can choose $M_{\rm in} = 1$. This however prevents us from studying the limit when we start from zero temperature, i.e. pure AdS background in the far past. Furthermore, the approximation of the delta function by a narrow Gaussian implies that we cannot study the case when $M_{\rm f}$ is too large as we will be interested in the region $\omega/ M_{\rm f} \approx 1$ too. Despite these limitations, we are equipped to explore a diverse range of qualitatively different behaviour of the time-dependence of $\rho(\omega, t_{\rm av}, \mathbf{k})$.

\subsection{The thermal spectral function}
\label{thermalspectralfunction}

Perhaps the simplest test of our prescription is to reproduce the well-known holographic thermal spectral function. Due to time translation invariance, $\rho(\omega, t_{\rm av}, \mathbf{k})$ does not depend on $t_{\rm av}$ in thermal equilibrium. The dual gravitational background is simply the background \eqref{AdS-Vaidya-gen} with $M(t) = M_0$, a constant. Thus this is just a special case of the discussion in the previous subsection.

It is useful to study
\begin{equation}
\rho_{\rm sbt} = \rho - \rho_{\rm AdS} =  \rho - \rho_{ T = 0},
\end{equation}
where $ \rho_{\rm AdS} = \rho_{ T = 0}$ is given by \eqref{rho-AdS}, in order to extract the purely thermal contribution. Furthermore, it is also useful to study 
\begin{equation}
\tilde\rho = \rho - 2\omega,
\end{equation}
since $2\omega$ is the ubiquitous state-independent contact term as evident from \eqref{relation-simple-3}. Note that this term cannot be removed by counterterms. When $\mathbf{k} = 0$, $\rho_{\rm AdS} = 2 \omega$. Therefore, $\rho_{\rm sbt} = \tilde\rho$ when $\mathbf{k} = 0$. We use $\sigma < 0.01$ in our computations.

\begin{figure}[t]
\centering
\subfigure[$\rho(\omega, \mathbf{k})$]{\includegraphics[height=4.5cm]{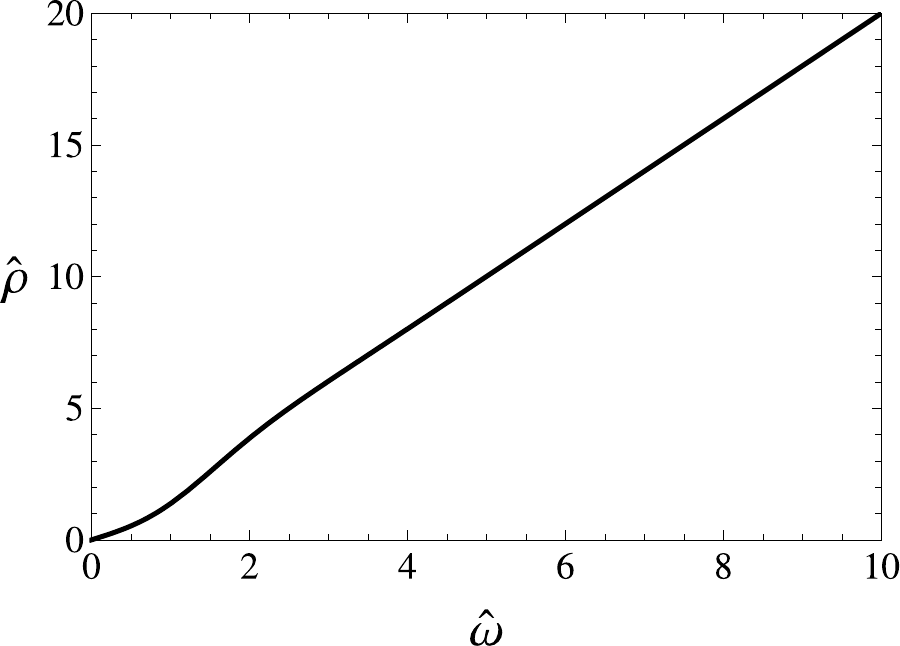}\label{fig:sch_rho_k0a}}
\subfigure[$\rho_\mathrm{sbt}(\omega, \mathbf{k})$]{\includegraphics[height=4.5cm]{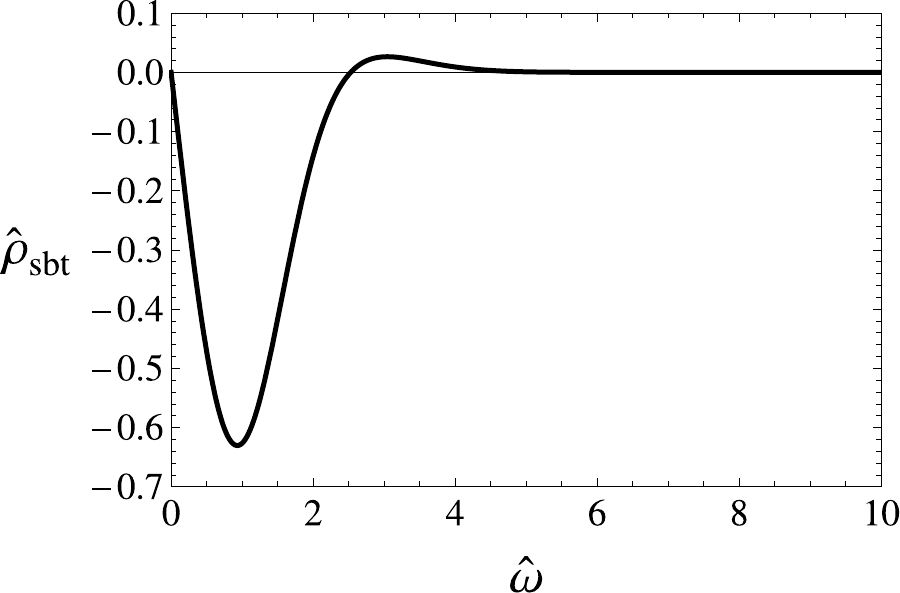}\label{fig:sch_rho_k0b}}
\caption{Thermal spectral function for $\mathbf{k}=0$.}
\label{fig:sch_rho_k0}
\end{figure}

\begin{figure}[t]
\centering
\subfigure[$\rho(\omega, \mathbf{k})$]{\includegraphics[height=4.5cm]{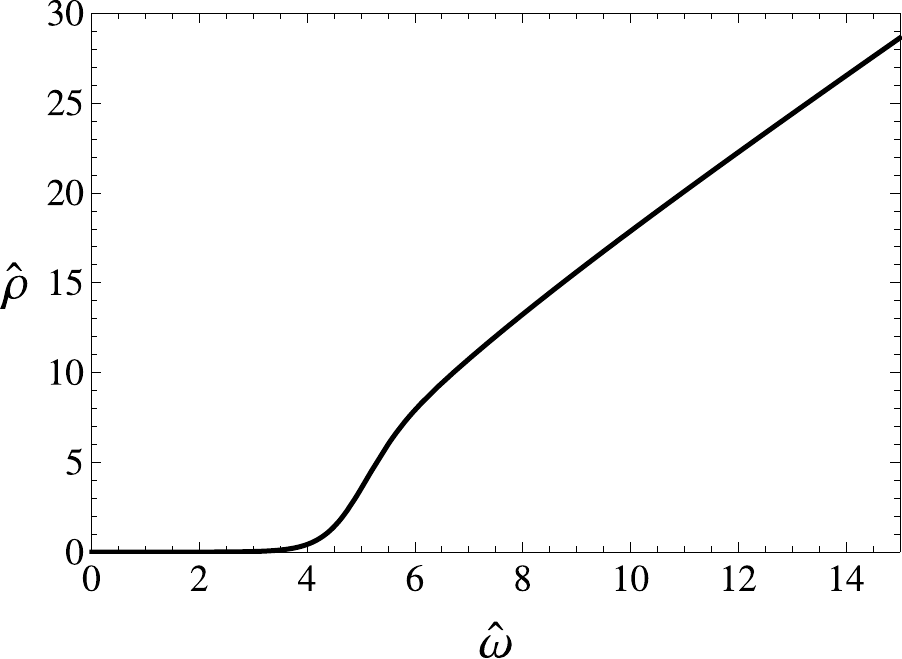}\label{fig:sch_rho_k4p5a}}
\subfigure[$\rho_\mathrm{sbt}(\omega, \mathbf{k})$]{\includegraphics[height=4.5cm]{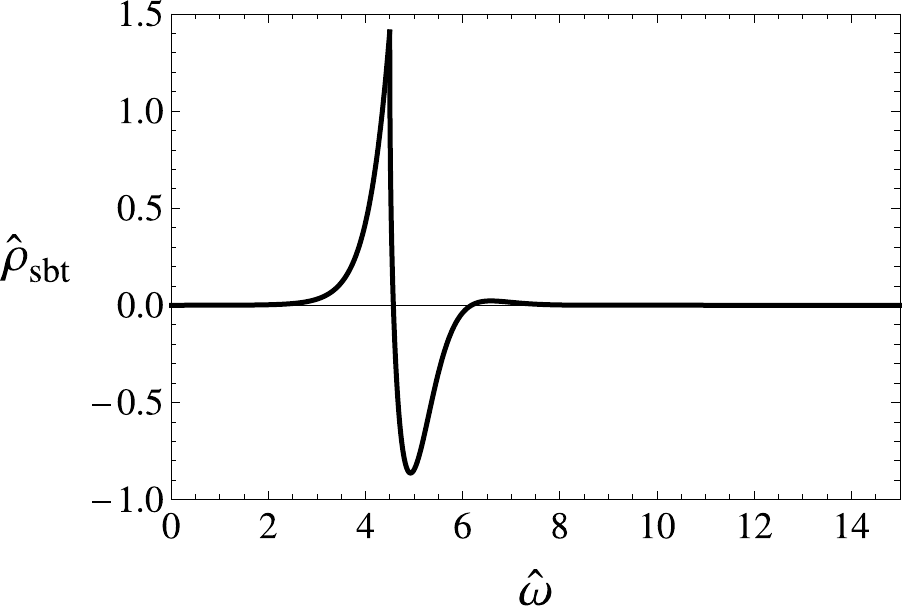}\label{fig:sch_rho_k4p5b}}
\subfigure[$\rho(\omega, \mathbf{k})-2\,\omega$]{\includegraphics[height=4cm]{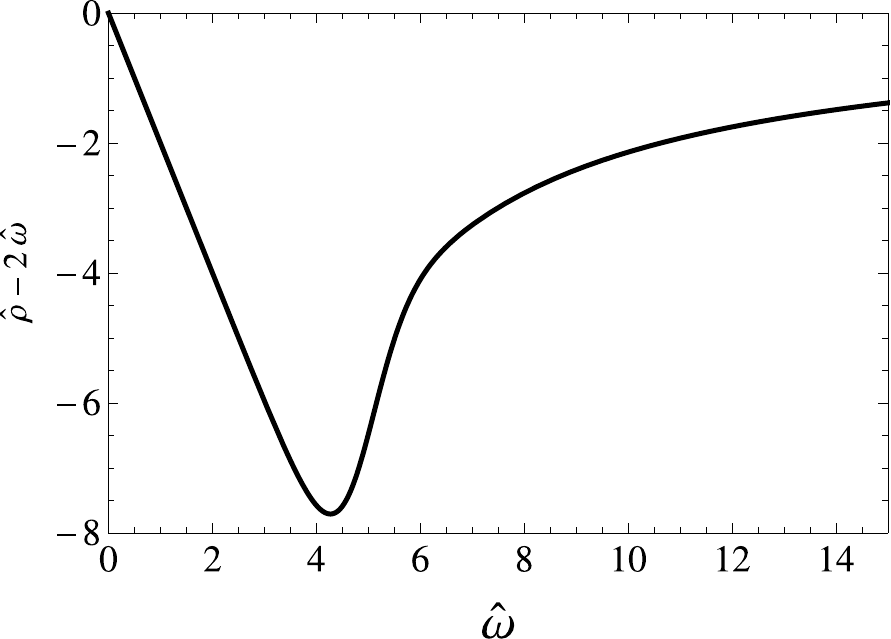}\label{fig:sch_rho_k4p5c}}
\caption{Thermal spectral function for $\hat{\mathbf{k}}=4.5$.}
\label{fig:sch_rho_k4p5}
\end{figure}

In Fig.~\ref{fig:sch_rho_k0}, the thermal spectral function for $\mathbf{k}=0$ obtained using our method is shown.  In the plots, we define a dimensionless quantities $\hat{X}$ from its dimensionful counterpart $X$ as $\hat{X} \equiv (3/4 \pi T_{\rm H})^nX$ with $n$ being the mass dimension of $X$. In Fig.~\ref{fig:sch_rho_k0a}, we plot the thermal spectral function, and shown in Fig.~\ref{fig:sch_rho_k0b} is $\rho_\mathrm{sbt}$ in which the linear growth in the large $\omega$ region gets subtracted away. Similarly, in Fig.~\ref{fig:sch_rho_k4p5} results for $\hat{\mathbf{k}}=4.5$ are shown. From Fig.~\ref{fig:sch_rho_k4p5a}, we find the spectrum is exponentially suppressed in $\omega<\mathbf{k}$. In Fig.~\ref{fig:sch_rho_k4p5b}, the function in $\omega<\mathbf{k}$ is the same as that in Fig.~\ref{fig:sch_rho_k4p5a}, but there is a sharp peak at $\omega=\vert \mathbf{k}\vert$ because of the subtraction. The spectral function in $\omega >\mathbf{k}$ has the qualitatively same distribution as in Fig.~\ref{fig:sch_rho_k0a}. Besides $\rho$ and $\rho_\mathrm{sbt}$, we also plot $\tilde{\rho}$ in Fig.~\ref{fig:sch_rho_k4p5c} which is determined by $f_1$ contribution as in \eqref{relation-simple-3}. For small $\omega$, the spectral function depends linearly on $\omega$ because of the $2\omega$ subtraction.

Using the well known method by Son and Starinets \cite{Son:2002sd} for obtaining the holographic thermal spectral function (see \cite{Teaney:2006nc,Kovtun:2006pf} for examples), we have checked that the plots in Figs.~\ref{fig:sch_rho_k0} and \ref{fig:sch_rho_k4p5} and similarly those for other values of $\vert \mathbf{k} \vert$ are reproduced. However, actual procedures in calculations are quite different,\footnote{The main difference between our method and that of Son and Starinets is that we solve an initial value problem where we impose regularity at the horizon via the choice of initial conditions, whereas in the latter case a boundary value problem is solved with the infalling boundary condition at the horizon.}
 and our method thus surely passes the simplest test of reproducing the holographic thermal spectral function.

\section{Results and a classification of routes to thermalisation}
\label{Results}

In the previous section we have discussed a general prescription for finding the time-dependent retarded propagator using holography in generic non-equilibrium states of a strongly coupled large $N$ theory, and have also shown that it reproduces the well-known thermal case. Furthermore, we have discussed how we can adapt our method to the specific AdS-Vaidya gravitational backgrounds (\eqref{AdS-Vaidya-gen} and \eqref{M-t}) that holographically capture features of spatially homogeneous non-equilibrium dynamics driven by a quench to a certain degree of approximation. Here we will present our numerical results and classify various routes to thermalisation of the time-dependent spectral function by varying the time duration of the quench. We will find various qualitative patterns of thermalisation of the spectral function as we interpolate between adiabatic and  instantaneous quench limits (these limits are described below). Our results will be specifically for a scalar operator with $\Delta =2$ in a $(2+1)-$dimensional holographic CFT -- in this case we will obtain exact results going beyond the geodesic approximation examined extensively in the literature.

\begin{figure}[!h]
\centering
\includegraphics[scale=0.55]{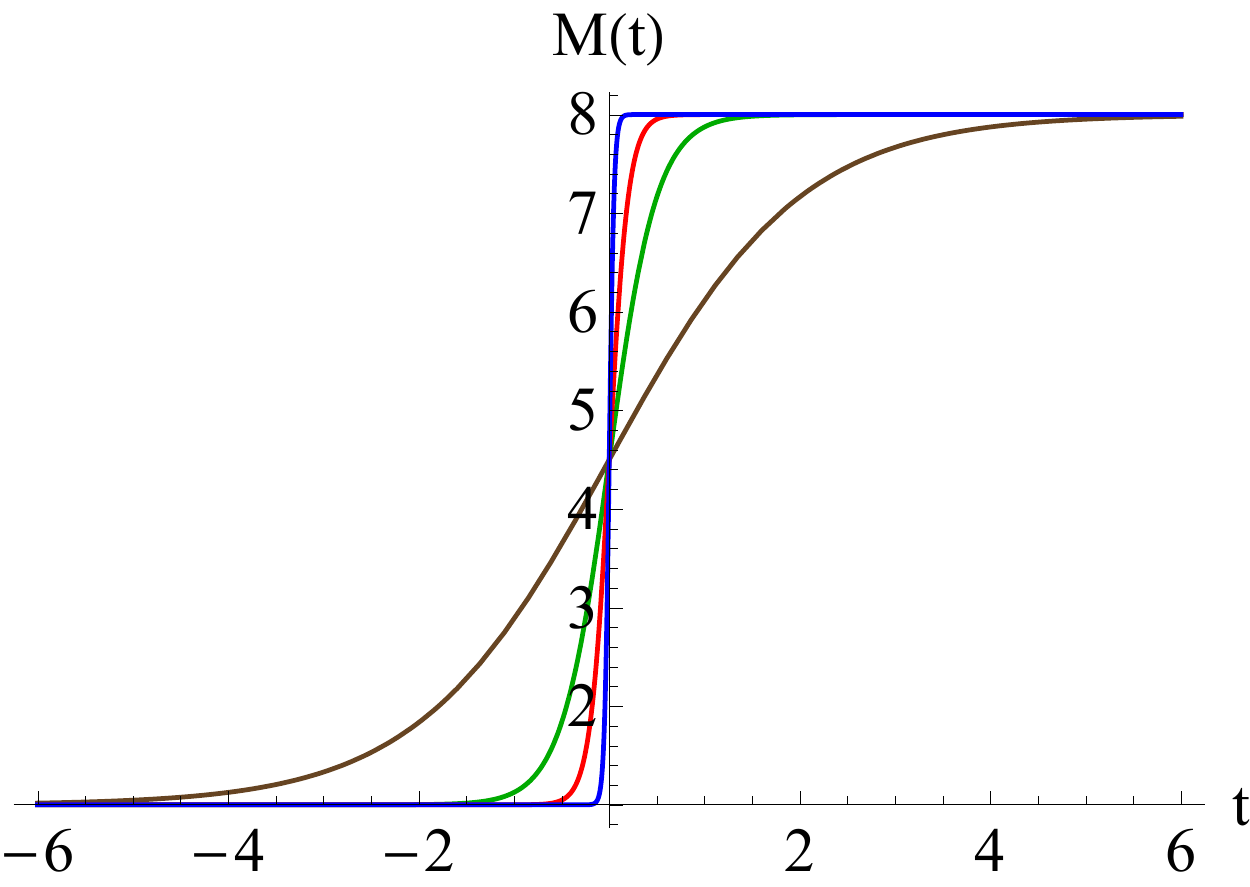}
\caption{Mass function for $\alpha=2.0$(brown), $0.5$(green), $0.2$(red) and $0.05$(blue)}
\label{fig:mass}
\end{figure}

The \textit{quenching parameter} $\alpha$ in \eqref{M-t} labeling the $(3+1)-$AdS-Vaidya gravitational backgrounds gives us the effective transition time from the initial to the final thermal equilibrium represented by AdS-Schwarzschild black holes of masses $M_{\rm in} = 1$ and $M_{\rm f} =8$ respectively (note $M_{\rm in} = 1$ also sets our units for field-theory observables. From now on we remove hats from the dimensionless quantities in the plots). The Bondi mass function $M(t)$ of these geometries are plotted in Fig. \ref{fig:mass} for various values of $\alpha$. Clearly $\alpha \rightarrow 0$ is the instantaneous quench limit and $\alpha \rightarrow \infty$ is the adiabatic limit. The mid-point of the transition is always chosen to be $t=0$. We may readily identify a time $t_{\rm i}^\alpha$ for a given $\alpha$ such that for $t < t_{\rm i}^\alpha$, the mass function $M(t)$ (or the effective instantaneous temperature) deviates from $M_{\rm in}$ (the initial temperature) by less than $1$ percent,  and similarly a time $t_{\rm f}^\alpha$ such that  for $t > t_{\rm f}^\alpha$ the mass function deviates from $M_{\rm f}$ (the final temperature) by less than $1$ percent.

\begin{figure}[t]
\centering
\subfigure[$\vert G_R(t,t_0; \mathbf{k} = 0) \vert$ when $\alpha = 0.05$.]{\includegraphics[height=5cm]{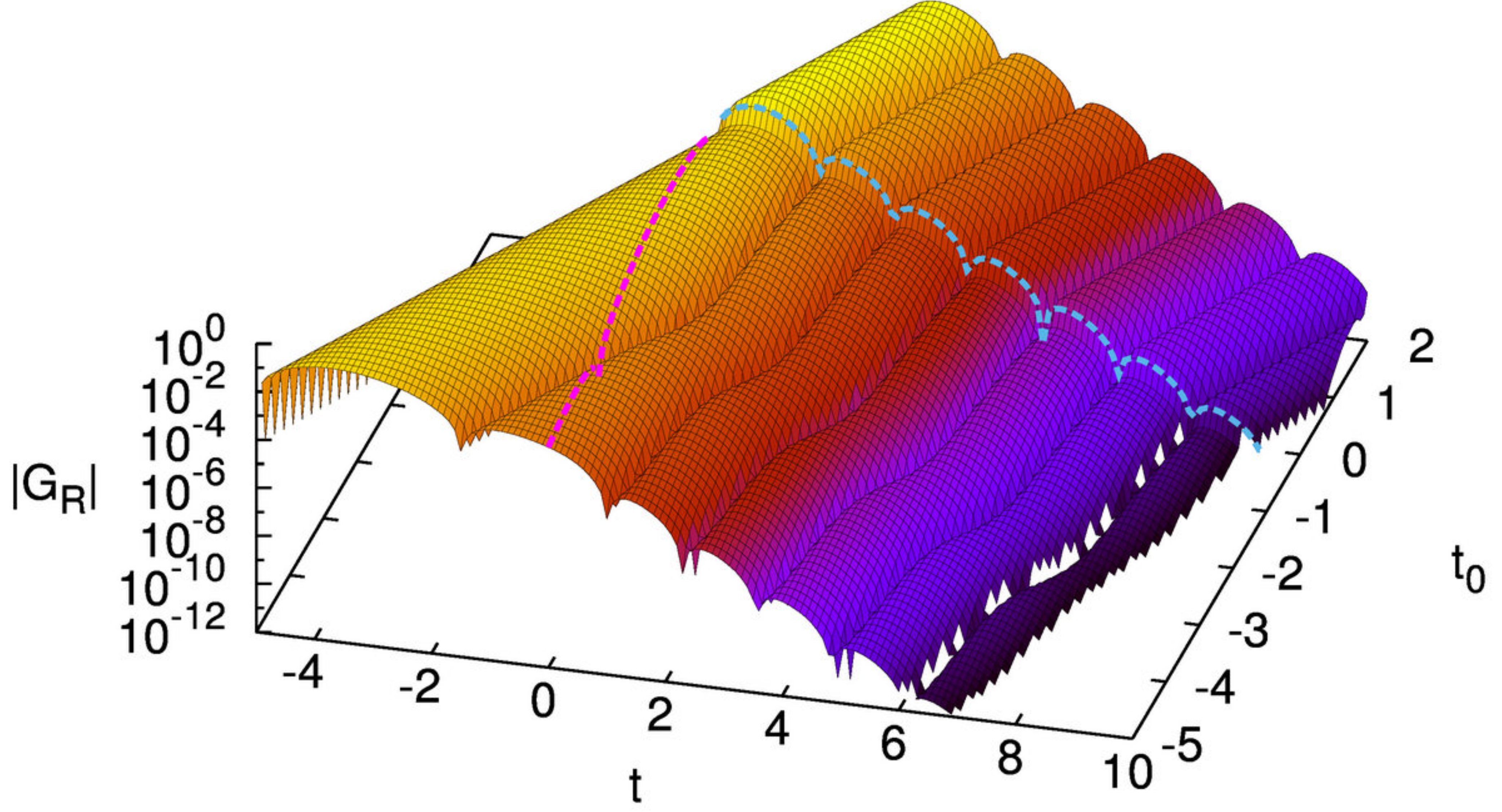}\label{fig:GR_k0a005_1}}
\subfigure[$\vert G_R(t,t_0;\mathbf{k}=0) \vert$ when $\alpha=2.0$.]{\includegraphics[height=5cm]{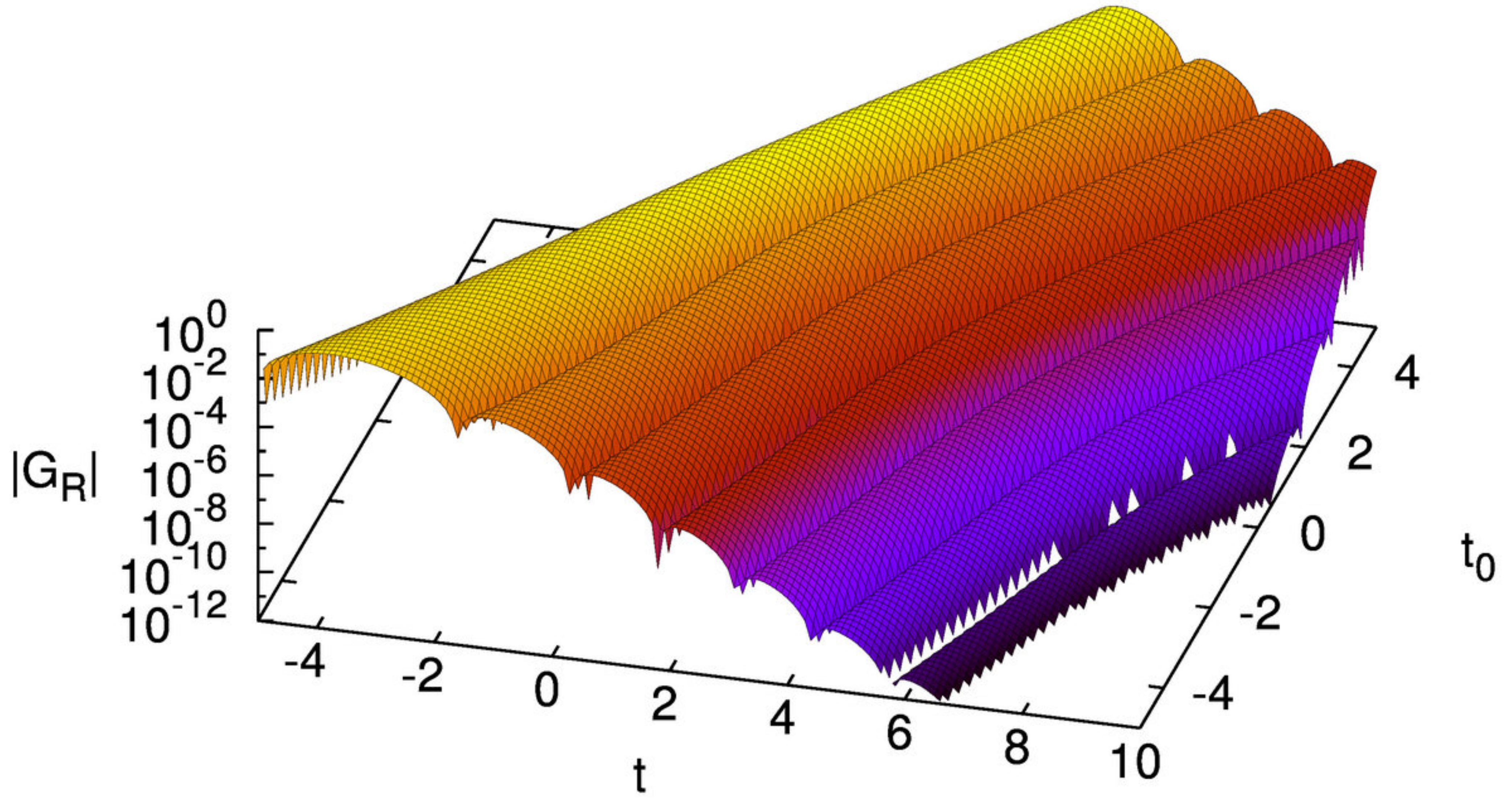}\label{fig:GR_k0a2}}
\caption{The contact term $\dot{\delta}(t- t_0)$ has been subtracted. For visual clarity, we have added dashed lines at $t=0$ and $t_0=0$ in Fig. (a).}
\label{fig:GR_k0a005}
\end{figure}

In Section \ref{example}, we have illustrated our method for computing $G_R^{OO} (t, t_0; \mathbf{k})$ (from now on we drop the superscripts $OO$). We present the three dimensional plots for $\vert G_R (t, t_0; \mathbf{k} = 0) \vert$ in case of rapid quench ($\alpha = 0.05$) and in case of slow quench ($\alpha = 2.0$) in Figs. \ref{fig:GR_k0a005_1} and  \ref{fig:GR_k0a2} respectively. The contact term $\dot{\delta}(t- t_0)$ in \eqref{relation-simple-2} has been subtracted in these plots, which therefore simply show $\vert f_1(t,\mathbf{k} = 0)\vert$ that results from causal response to the source $f_0(t',\mathbf{k} = 0) = \delta (t' - t_0)$. Clearly, we see that $G_R$ vanishes for $t < t_0$ as expected from causality. Across $t_0 = 0$, when the background transits from a black brane with temperature $T_{\rm in}$ to that with temperature $T_{\rm f}$, the decay and oscillation rates of $\vert G_R \vert$ change by a factor of $T_\mathrm{f}/T_\mathrm{in} = 2$ as we should expect from quasi-normal mode analysis.  Also the \textit{wake up} of $\vert G_R \vert$ at $t = t_0$ gets enhanced by a factor of $(T_\mathrm{f}/T_\mathrm{in})^2 = 4$ as we move from $t_0 = - \infty$ to $t_0 = \infty$. These transitions of decay and oscillation rates, and also wake up amplitudes are rapid in the case of $\alpha = 0.05$, but slow when $\alpha = 2.0$.


Our strategy for finding $\rho(\omega, t_{\rm av},\mathbf{k})$  has also been discussed in detail in Section \ref{example}. We will actually study the subtracted spectral function
\be
\label{tilderho}
{\tilde\rho} (\omega, t_{\rm av}, \mathbf{k})= \rho (\omega, t_{\rm av}, \mathbf{k})-2 \omega,
\ee 
with the state-independent contact term $2\omega$ removed. In order to evaluate ${\tilde\rho} (\omega, t_{\rm av}, \mathbf{k})$ therefore, we just need to perform the integral in \eqref{relation-simple-3} for various values of $t_{\rm av}$. This is done in two steps. First, for a fixed $t_{\rm av}$ we obtain the response $f_1(t,\mathbf{k})$ for the source effectively $f_0(t') = \delta(t' -t_0)$ with $t_0 = 2t_{\rm av}-t$. Both $t$ and $t_0$ are varied keeping $t_{\rm av}$ fixed. Note causality implies that $f_1(t,\mathbf{k})$ vanishes for $t< t_0$. Furthermore, if $t$ and $t_0$ are far apart, i.e. if $t_{\rm rel} = t -t_0$ is large, $f_1(t,\mathbf{k})$ decays exponentially at a rate controlled by the final black hole temperature -- so the integral does not receive significant contribution for large values of $t_{\rm rel}$. In the second step, the variable $t_{\rm rel} = t -t_0$ is replaced by the variable $t$ so that the integration in \eqref{relation-simple-3} can be performed numerically by discrete summation over $f_1(t, \mathbf{k})$ for various values of $t$ (we use step size of 0.01). We thus obtain ${\tilde\rho} (\omega, t_{\rm av}, \mathbf{k})$ as a function of $\omega$ for various values of $t_{\rm av}$ at a fixed value of $\vert \mathbf{k} \vert$.

Clearly for any value of $\alpha$, the spectral function $\rho(\omega, t_{\rm av}, \mathbf{k})$ will coincide with that of the thermal form corresponding to the initial black hole as $t_{\rm av} \rightarrow -\infty$ and with that of the thermal form corresponding to the final black hole as $t_{\rm av} \rightarrow \infty$. Nevertheless, the effective transition time will depend on $\alpha$. We denote the time $t_{\rm av(i)}^\alpha$ as that for which $\rho(\omega, t_{\rm av}, \mathbf{k})$ becomes independent of $t_{\rm av}$ when $t_{\rm av} < t_{\rm av(i)}^\alpha$ and coincides with the thermal form of the initial black hole up to 1 percent numerical difference\footnote{Note again that we are always referring to dimensionless quantities in units where $M_{\rm in} = 1$.} (to be precise we mean that the maximal deviation is less than 1 percent). Similarly we define $t_{\rm av(f)}^\alpha$ such that for $t > t_{\rm av(f)}^\alpha$, $\rho(\omega, t_{\rm av}, \mathbf{k})$ becomes independent of $t_{\rm av}$ coinciding with the thermal form of the final black hole up to 1 percent numerical difference (to be precise we mean that the maximal deviation is less than 1 percent again). Interestingly, $ t_{\rm av(i)}^\alpha$ and $t_{\rm av(f)}^\alpha$ need not coincide with $t_{\rm i}^\alpha$ and $t_{\rm f}^\alpha$ respectively, where the mass function $M(t)$ coincides with the initial and final values up to 1 percent numerical difference. Indeed, this turns out to be true specially when $\alpha$ is small, $\alpha \le 0.1$ and very large $\alpha \ge 2$ as shown below.

We divide our study in three broad categories, namely,  $\vert \mathbf{k}\vert < M_{\rm in}$, $\vert \mathbf{k}\vert \sim M_{\rm in}$  and $\vert \mathbf{k}\vert > M_{\rm in}$. The first category is represented by the choice $\vert \mathbf{k}\vert = 0$, the second one by $\vert \mathbf{k}\vert = 1$ and the third by $\vert \mathbf{k}\vert = 4.5$. The regime $\vert \mathbf{k}\vert < M_{\rm in}$ for $M_{\rm in} > 0$ particularly shows qualitatively new patterns of thermalisation distinct from what has been found in the literature before.

\begin{table}[h]
 \begin{tabular}{| p{1.1cm} | p{2.3cm} | p{2.3cm}| p{2.3cm}|  p{2.3cm} |p{2.3cm} |}
      \hline
    $\alpha$ & Delayed/ Advanced time-dependence & Kink formation & Persistent oscillations &  Mild/ large distortions & Slightly premature thermalisation \\ \hline\hline
    0.05 & Advanced & Present & Present & Large & Absent \\ \hline
     0.2 & Advanced but suppressed  & Present & Suppressed & Large & Absent \\ \hline
     0.5 & Absent & Absent & Absent & Mild & Absent \\ \hline
     2.0 & Delayed & Absent & Absent & Absent & Present \\ \hline
    \end{tabular}
\caption{Patterns of thermalisation at various values of $\alpha$ for $|\mathbf{k}|  = 0$}\label{k=0}
 \end{table}
 \begin{table}[h]
 \begin{tabular}{| p{1.1cm} | p{2.3cm} | p{2.3cm}| p{2.3cm}|  p{2.3cm} |p{2.3cm} |}
      \hline
    $\alpha$ & Delayed/ Advanced time-dependence & Kink formation & Persistent oscillations &  Mild/ large distortions & Slightly premature thermalisation \\ \hline\hline
    0.05 & Advanced & Present & Present & Large & Absent \\ \hline
     0.2 & Advanced but suppressed & Suppressed & Suppressed & Large & Absent \\ \hline
     0.5 & Absent & Absent & Absent & Mild & Absent \\ \hline
     2.0 & Delayed & Absent & Absent & Absent & Present \\ \hline
    \end{tabular}
\caption{Patterns of thermalisation at various values of $\alpha$ for $|\mathbf{k}|  = 1$}\label{k=1}
 \end{table}
\begin{table}[h]
 \begin{tabular}{| p{1.1cm} | p{2.3cm} | p{2.3cm}| p{2.3cm}|  p{2.3cm} |p{2.3cm} |}
      \hline
    $\alpha$ & Delayed/ Advanced time-dependence & Kink formation & Persistent oscillations &  Mild/ large distortions & Slightly premature thermalisation \\ \hline\hline
    0.05 & Advanced & Suppressed & Suppressed & Mild & Absent \\ \hline
     0.2 & Advanced but suppressed & Absent & Suppressed & Mild & Absent \\ \hline
     0.5 & Absent & Absent & Absent & Absent& Absent \\ \hline
     2.0 & Delayed & Absent & Absent & Absent & Present \\ \hline
    \end{tabular}
\caption{Patterns of thermalisation at various values of $\alpha$ for $|\mathbf{k}|  = 4.5$}\label{k=4.5}
 \end{table}

Tables \ref{k=0}, \ref{k=1} and \ref{k=4.5} summarise the various routes of thermalisation at a glance for appropriately chosen values of $\alpha$ and $\vert \mathbf{k} \vert$ by specifying the presence/absence or degrees of appearance of various possible features that are going to be described below. We find distinct patterns of thermalisation developing as we increase $\alpha$ and $\vert \mathbf{k} \vert$. Each pattern will be discussed at length in the following subsections. It is to be noted that the transitions between these patterns as we increase $\alpha$ and $\vert \mathbf{k} \vert$ are not very sharp. We have also not attempted to specify the boundaries of $\alpha$ at fixed $\vert \mathbf{k} \vert$ where each pattern is observed even approximately, however we do find smooth transitions from one pattern to another as we vary $\alpha$ and $\vert \mathbf{k} \vert$.

\subsection{Numerical Results} 
\noindent
In what follows, we will present our numerical plots for the time evolution of the spectral function starting with the lowest momentum $\vert \mathbf{k}\vert=0$ and going all the way to $\vert \mathbf{k}\vert=4.5$. In order to make the dynamics of thermalisation visible from the figures, we have chosen the following colour codes:  green when the spectral function coincides with the thermal form of the initial black hole at $t_{\rm av}  = t_{\rm av(i)}$, red when it coincides with the thermal form of the final black hole at $t_{\rm av}  = t_{\rm av(f)}$ and the other colours are used in the window $t_{\rm av(i)}\le t_{\rm av}\le t_{\rm av(f)}$. We will consider three different values of $\vert \mathbf{k}\vert$ representing three different cases, namely $\vert \mathbf{k}\vert < M_{\rm in}$, $\vert \mathbf{k}\vert \sim M_{\rm in}$ and $\vert \mathbf{k}\vert > M_{\rm in}$ respectively.

\subsubsection{$\vert \mathbf{k} \vert=0$ representing $\vert \mathbf{k} \vert < M_{\rm in}$}
As shown in Table \ref{k=0}, this regime is the richest with four distinct routes of thermalisation each corresponding to a different value of $\alpha$, namely $0.05$, $0.2$, $0.5$ and $2$.
We study these cases so that we we can interpolate from instantaneous quench to adiabatic transition in the dual non-equilibrium state. Although we examine them for $\vert \mathbf{k} \vert=0$, these also hold when $\vert \mathbf{k} \vert \ll M_{\rm in}$, as for instance when $\vert \mathbf{k} \vert=0.5$.  At early and late times, the spectral function has peaks corresponding to the thermal quasinormal mode of the initial/final black holes. The location and width of these peaks are proportional to the temperature. In all the four patterns of thermalisation, the width of the spectral peak broadens uniformly reflecting rise in the effective temperature (note the mass function $M(t)$ increases monotonically). Nevertheless, there can be various forms of distortions.

\begin{figure}[!h]
\includegraphics[scale=0.9]{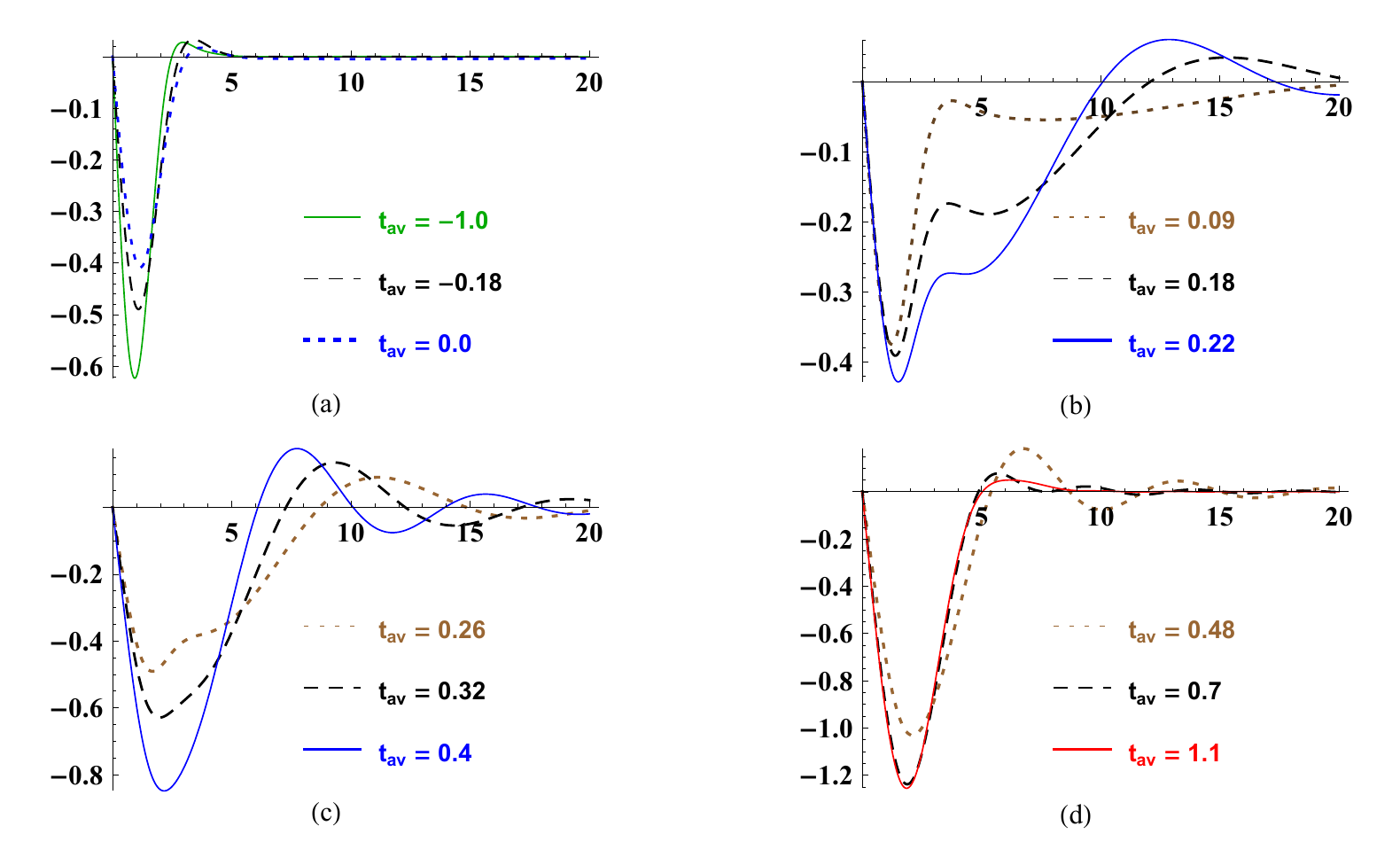}
\caption{$\tilde{\rho}$ as a function of $\omega$ for $\vert \mathbf{k}\vert=0$ and $\alpha=0.05$. Here, $t_i \sim -0.17$, $t_f \sim 0.17$, $t_{\rm av(i)} \sim -1.0$, $t_{\rm av (f)} \sim 1.1$ }
\label{fig:pna1}
\end{figure}

\textbf{(a) $\mathbf{\alpha=0.05}$:} The most dramatic route to thermalisation is obtained for $\alpha=0.05$. The plots are in Fig \ref{fig:pna1}.  
There are three distinct features. 
\begin{enumerate}
\item \textbf{Advanced time-dependence:} The spectral function starts getting time-dependent much before $t_{\rm av} =t_{\rm i}^\alpha$, which is when the mass function (effective temperature) picks up a significant change -- i.e. $t_{\rm av(i)}^\alpha < t_{\rm i}^\alpha$. This is not in contradiction with causality because the spectral function is non-local in time by definition. However, this advanced time-dependence is pronounced only for low values of $\alpha$. Here $t_{\rm av(i)}^\alpha = -1.0$ whereas $t_{\rm i}^\alpha= -0.17$. A similar advanced time-dependence has been observed in \cite{Keranen:2015mqc} in the fast quench limit. It is worth noting at this point that for large values of $\alpha$ corresponding to the adiabatic limit, we will find the opposite behaviour, namely \textit{delayed acceleration}, i.e. $t_{\rm av(i)}^\alpha > t_{\rm i}^\alpha$.
 
\item \textbf{Kinkiness:} Around the time $t_{\rm av}=0$,  the spectral peak starts getting distorted
for $\omega > M_{\rm in}$ with significant spectral weight transferred to the higher frequencies. At $t_{\rm av}=0.18, 0.22$ and $0.26$, we can see a sharp kink in the spectral function in the high frequency region. This feature of large spectral weight transfer to higher frequencies $\omega > M_{\rm in}$ will also be present for somewhat higher values of $\alpha$. More precisely, the spectral weight transfer to higher frequencies is defined here in comparison with the form of the thermalised spectral function at the instantaneous effective temperature.
\item \textbf{Persistent oscillations:} For higher values of $t_{\rm av}$, the kinks disappear, but smaller peaks start appearing and disappearing thus forming transient ripples before the spectral function settles down to the final thermal form at $t_{\rm av(f)}$. As visible in Fig.~\ref{fig:pna1}(c), the ripples propagate towards lower frequencies. The amplitude of these ripples decrease with time, but the number of visible peaks increase (indeed zooming inside each peak we find secondary peaks as well -- this is however not shown in the plots).\footnote{We speculate that a fractal structure may form at lower values of $\alpha$. This may be related to the non-analyticity at the onset of thermalisation found in \cite{Balasubramanian:2011ur} with the geodesic approximation. In order to explore lower values of $\alpha$, we need to decrease our time-step. This translates to increase in computational time. In future work we will investigate if it is indeed feasible to take the instantaneous quench limit continuously within our approach.}  As is clear, these oscillations persist long after $t_{\rm av} = t_{\rm f}^\alpha$ so that $t_{\rm (av)f}^\alpha \gg t_{\rm f}^\alpha$. 
\end{enumerate}
 
Note that the height of the spectral function peak first decreases as we increase the time from $t_{\rm av(i)}$. After some time, $t_{\rm av} \approx 0.09$, the peak height starts increasing with time until the spectral function attains its final form. 

We also did the numerics for $\alpha=0.1$ and the qualitative features remain exactly the same as that for $\alpha=0.05$.

It is to be noted though that the advanced time dependence and persistent oscillations are not mere artefacts of the Fourier transform involved in the definition of the Wigner transform of the commutator. In fact, these will be absent as we increase the duration of the quench, for larger values of which advanced time-dependence will be replaced by delayed time-dependence.

\begin{figure}[!h]
\hspace{0.1cm}
\includegraphics[scale=0.95]{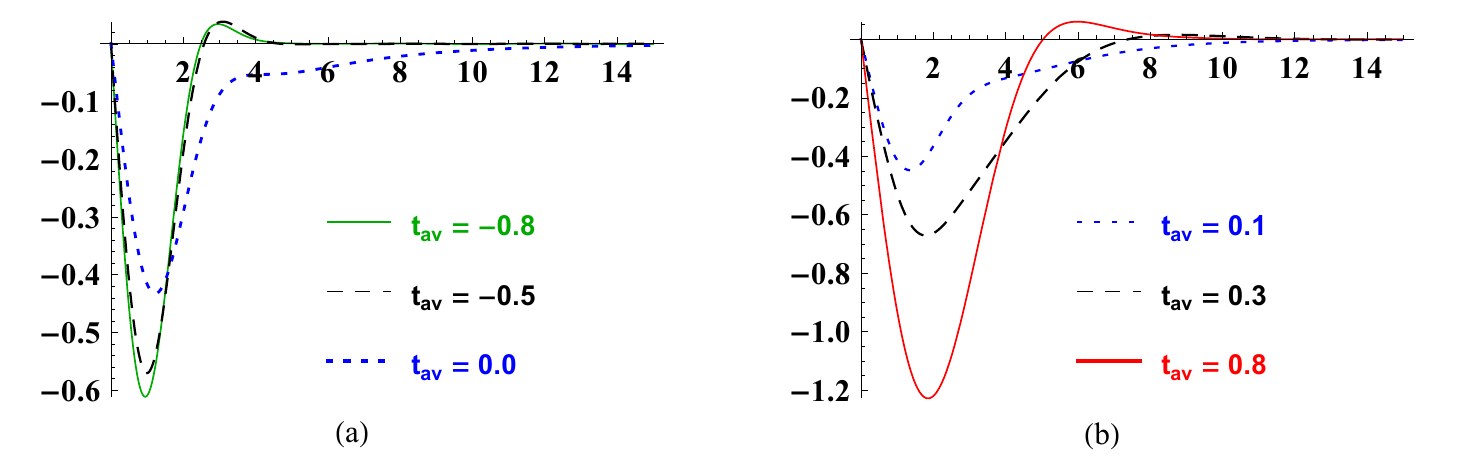}
\caption{$\tilde{\rho}$ as a function of  $\omega$ for $\vert \mathbf{k}\vert=0$ and $\alpha=0.2$. Here, $t_i \sim -0.66$, $t_f\sim 0.66$, $t_{\rm av(i)}\sim -0.8$, $t_{\rm av (f)}\sim 0.8$. }
\label{fig:kna}
\end{figure}

\textbf{(b) $\mathbf{\alpha=0.2}$:} The pattern of thermalisation found for $\alpha=0.2$ is depicted in Fig \ref{fig:kna}. Here, we find that the advanced time-dependence is suppressed -- in this case $t_{\rm i}^\alpha \approx -0.66$ and $t_{\rm av(i)}^\alpha \approx -0.8$.  Around $t_{\rm av} = 0$, the spectral function develops kinks involving the transfer of spectral weight towards higher frequency modes $\omega \gg M_{\rm in}$. Also note that for $t_{\rm av} > t_{\rm av(i)}$, the height of the spectral peak initially decreases, but then it starts increasing monotonically again as the spectral function assumes the final thermal form. The spectral function still takes slightly longer time to thermalise than the background mass function $M(t)$, i.e.,  $t_{\rm av(f)}^\alpha > t_{\rm f}^\alpha$ -- but the relative difference between the two is much smaller than that in the previous case. 

\begin{figure}[t]
\hspace{0.1cm}
\vskip -0.4cm
\includegraphics[scale=0.95]{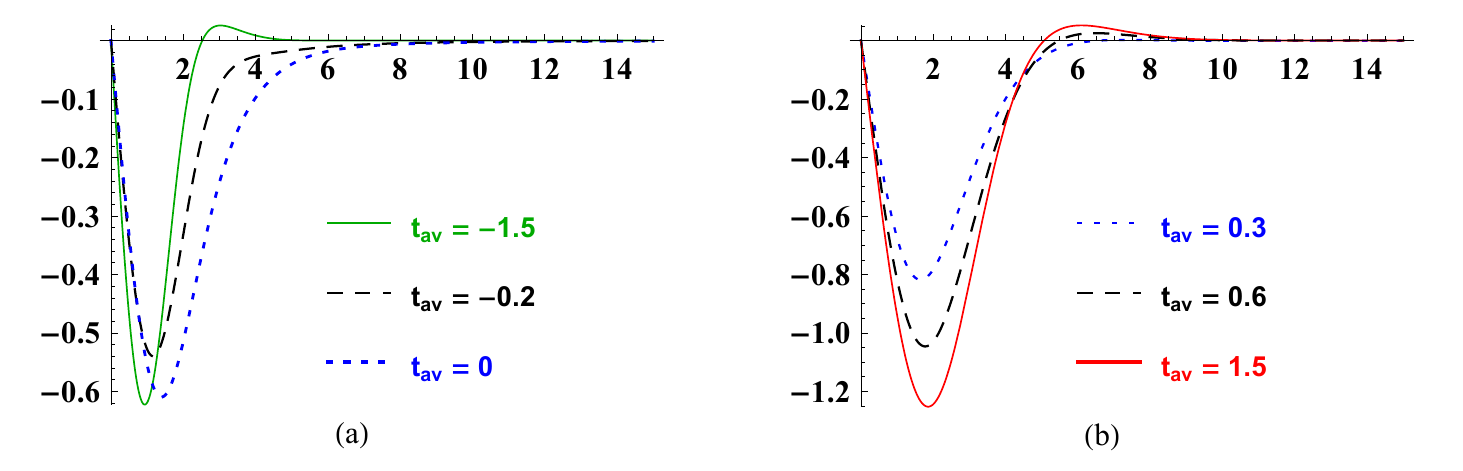}
\vskip -0.3cm
\caption{$\tilde{\rho}$ as a function of $\omega$ for $\vert \mathbf{k}\vert=0$ and $\alpha=0.5$. Here, $t_i \sim -1.55$, $t_f \sim 1.55$, $t_{\rm av(i)}\sim -1.5$, $t_{\rm av (f)}\sim 1.5$ }
\label{fig:mna}
\end{figure}

\textbf{(c) $\mathbf{\alpha=0.5}$:} 
Going to a slightly longer duration of the quench by choosing $\alpha=0.5$, we find that the thermalisation remains 
non-adiabatic but the qualitative features are very different from the previous two cases as shown in Fig.
\ref{fig:mna}. There is no significant advanced time dependence. However, we find a very tiny transfer of spectral weight around $t_{\rm av}=0$ and we 
call such a thermalisation as mildly non-adiabatic because there are only \textit{mild distortions} in comparison with the form of the thermalised spectral functions at the instantaneous effective temperatures. As in the previous case, the peak height decreases first a bit before it starts increasing monotonically again. The time scale of thermalisation of the spectral function to the final form is almost same as that for the mass function, i.e. $t_{\rm av(f)}^\alpha \approx t_{\rm f}^\alpha$.

\begin{figure}
\centering
\includegraphics[scale=1]{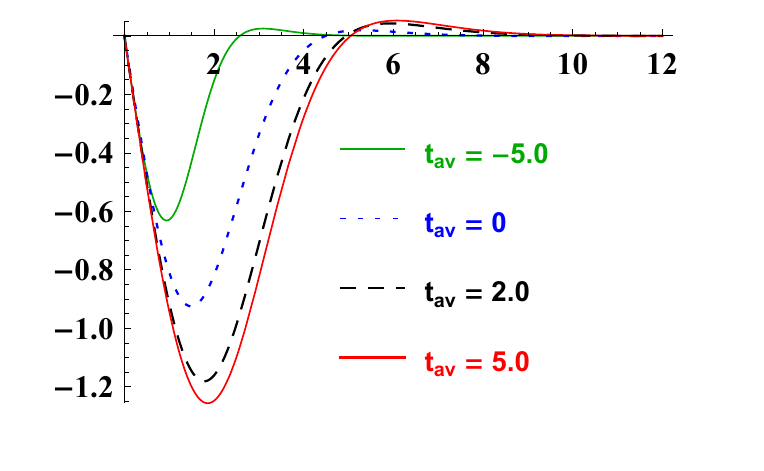}
\caption{$\tilde{\rho}$ as a function of $\omega$ for $\vert \mathbf{k}\vert=0$ and $\alpha=2.0$. Here, $t_i \sim -6.6$, $t_f\sim 6.6$, $t_{\rm av(i)}\sim -5.0$, $t_{\rm av (f)}\sim 5.0$. }
\label{fig:a}
\end{figure}

\textbf{(d) $\mathbf{\alpha=2.0}$:} From  the trends discussed above, it can be expected that tuning $\alpha$ to even higher values will lead to an adiabatic thermalisation. This indeed shows up in Fig. \ref{fig:a} for $\alpha=2.0$. We find that \textit{both} the height and width of the spectral peak monotonically increases. Due to the slow rate of change of background mass function, the spectral function approximately assumes the thermal form at the instantaneous effective temperature. However, the notion of instantaneous effective temperature is strictly not valid here as the mass is a function of time. Therefore, adiabatic thermalisation more precisely stands for uniform deformation of the spectral function without distortion.

There are two distinct features worth noting nevertheless: 
\begin{enumerate}

\item \textbf{Delayed time-dependence:} Significant time dependence (by 1 percent) in the spectral function commences with a slight delay from that for the mass function $M(t)$. Comparing Fig. \ref{fig:a} with Fig. \ref{fig:mass}, one can clearly see that indeed $t_{\rm av(i)}^\alpha(\approx -5)$ > $t_{\rm i}^\alpha (\approx -6.6)$. This feature has been found in the geodesic approximation before \cite{Balasubramanian:2011ur}. 

\item \textbf{Slightly premature thermalisation:} Furthermore, the final thermal form of the spectral function is attained (with 1 percent maximal deviation) slightly \textit{before} the time-dependence of the mass function becomes insignificant, i.e. $t_{\rm av(f)}^\alpha(\approx 5) < t_{\rm f}^\alpha (\approx 6.6)$, as also evident from Figs. \ref{fig:a} and \ref{fig:mass}.
\end{enumerate}

We find that the qualitative features of spectral function thermalisation for $|\mathbf{k}|=0.5$ are the same as that for $|\mathbf{k}|=0$ for the same values of $\alpha$. We present the plots in Appendix \ref{appendixA} for completeness. 

\subsubsection{$\vert \mathbf{k}\vert=1.0$ representing $\vert \mathbf{k} \vert \sim M_{\rm in}$}

Table \ref{k=1} summarises the patterns of thermalisation for $|\mathbf{k}|=1=M_{\rm in}$ for the same set of values for $\alpha$ as in the previous case. We find lesser spectral weight transfer to UV modes as compared to the transfer in the cases for $|\mathbf{k}|=0$ and $|\mathbf{k}|=0.5$ at lower values of $\alpha$. 
Even though some distortions are noted in the curves, kink formations and persistent oscillations are relatively suppressed. The plots in Fig. \ref{fig:k31} for $\alpha=0.05$ are
almost similar to those in Fig. \ref{fig:pna1} nevertheless. However, we find that the effective time for thermalisation of the spectral function, i.e. $t_{\rm av(f)}^\alpha$, decreases in comparison to $|\mathbf{k}|=0$ situation.
 Comparing time values on Figs. \ref{fig:pna1}(d) and \ref{fig:k31}(b), we readily note that $t_{\rm av(f)}^\alpha(|\mathbf{k}|=0) > t_{\rm av(f)}^\alpha(|\mathbf{k}|=1.0)$ for $\alpha = 0.05$. This implies that giving  momentum to the scalar field leads to decrease of the spectral function thermalisation time. 

At this point, we do see that the thermalisation of the spectral function has a very distinct behaviour compared to that of the one point function. It is known that the imaginary part of the quasinormal-mode frequency decreases as we increase $\vert \mathbf{k}\vert$ (see \cite{Starinets:2002br}), indicating that the one-point function takes longer to thermalise as we increase $\vert \mathbf{k}\vert$. We comment on this behaviour in the next subsection.

\begin{figure}[t]
\hspace{0.8cm}
\includegraphics[scale=0.95]{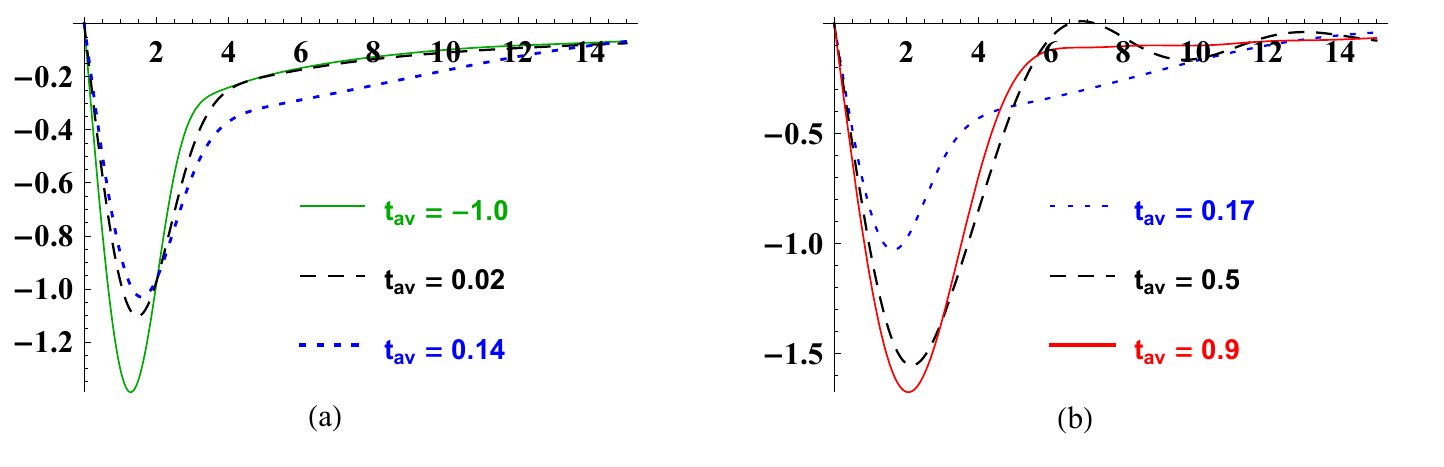}
\caption{$\tilde{\rho}$ as a function of $\omega$ for $\vert \mathbf{k}\vert=1$ and $\alpha=0.05$. Here, $t_i \sim -0.17$, $t_f \sim 0.17$, $t_{\rm av(i)} \sim -1.0$, $t_{ \rm av (f)} \sim 0.9$ } \label{fig:k31}
\label{fig:k31}
\end{figure}

\begin{figure}[t]
\begin{minipage}{.45\textwidth}
\hspace{0.7cm}\includegraphics[scale=.95]{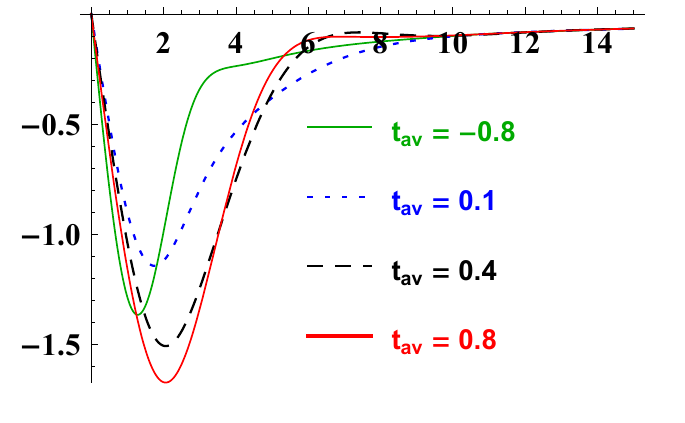}
 \caption{$\tilde{\rho}$ as a function of $\omega$ for $\vert \mathbf{k}\vert=1$ and $\alpha=0.2$. Here, $t_i \sim -0.66$, $t_f \sim 0.66$, $t_{\rm av(i)} \sim -0.8$, $t_{ \rm av (f)} \sim 0.8$ }
\label{fig:k33}
\end{minipage}
\begin{minipage}{0.1\textwidth}
\end{minipage}
\begin{minipage}{.45\textwidth}
\hspace{1cm} \includegraphics[scale=.95]{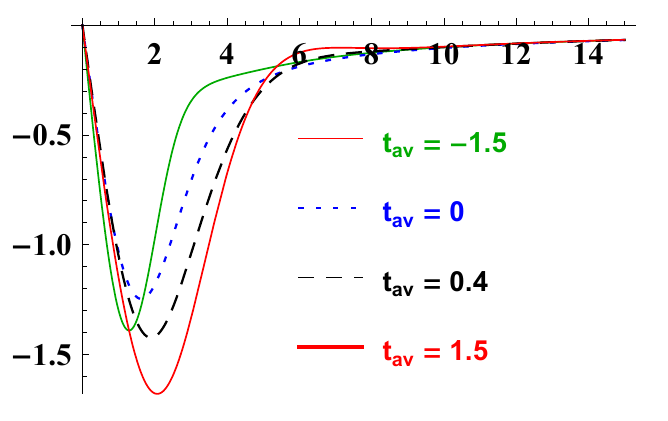}

   \caption{$\tilde{\rho}$ as a function of $\omega$ for $\vert \mathbf{k}\vert=1$ and $\alpha=0.5$. Here, $t_i \sim -1.55$, $t_f \sim 1.55$, $t_{\rm av(i)} \sim -1.5$, $t_{\rm av (f)} \sim 1.5$ }
  \label{fig:k34}
\end{minipage}
\end{figure}

\begin{figure}[t]
\hspace{4cm}\includegraphics[scale=0.95]{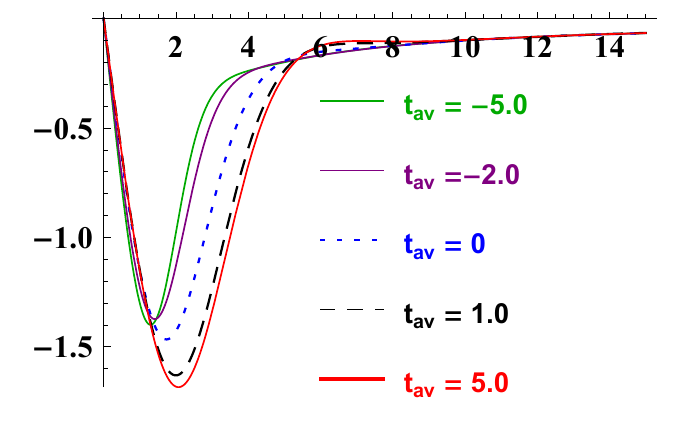}
\caption{$\tilde{\rho}$ as a function of $\omega$ for $\vert \mathbf{k} \vert=1$ and $\alpha=2.0$. Here, $t_i \sim -6.6$, $t_f \sim 6.6$, $t_{\rm av(i)} \sim -5.0$, $t_{\rm av (f)} \sim 5.0$  }
\label{fig:k35}
\end{figure}

As we increase the quenching parameter to $\alpha= 0.2$ and $0.5$, we find that mildly non-adiabatic features show up with the absence of prominent kink formation as shown in Figs.~\ref{fig:k33}, \ref{fig:k34}.
Similar to the zero momentum thermalisation patterns, we find the adiabatic (distortion-free) route at $\alpha=2$, this is shown in Fig. \ref{fig:k35}. However, note that in the adiabatic limit the height of the spectral peak initially decreases as we go above $t_{\rm av(i)}$ unlike the zero momentum case. Afterwards, the height increases monotonically. This is actually consistent with the behaviour of the thermal spectral functions at the instantaneous effective temperatures. Both delayed time-dependence and slightly premature thermalisation appear in the adiabatic limit. 

\subsubsection{$\vert \mathbf{k} \vert =4.5$ representing $\vert \mathbf{k} \vert > M_{\rm in}$}

\begin{figure}[t]
\centering
\includegraphics[scale=0.9]{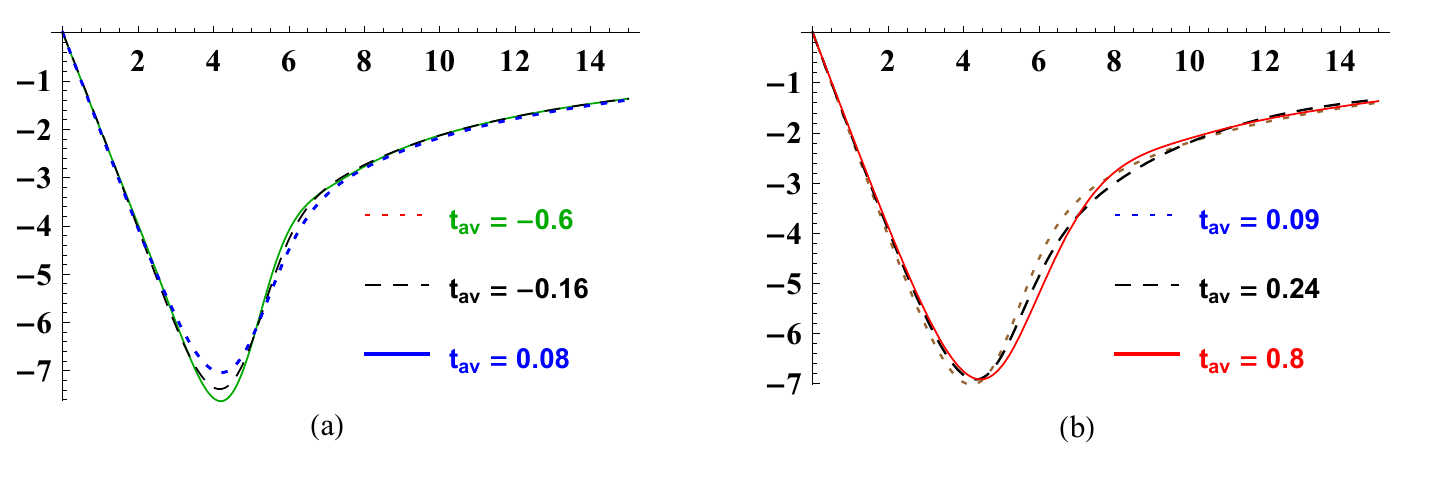}
\caption{$\tilde{\rho}$ as a function of $\omega$ for  $\vert \mathbf{k} \vert=4.5$ and $\alpha=.05$. Here, $t_i \sim -0.17$, $t_f \sim 0.17$, $t_{\rm av(i)} \sim -0.6$, $t_{ \rm av (f)} \sim 0.8$ }
\label{fig:a4}
\end{figure}

\begin{figure}[t]
\begin{minipage}{.45\textwidth}
\hspace{0.8cm}\includegraphics[scale=0.95]{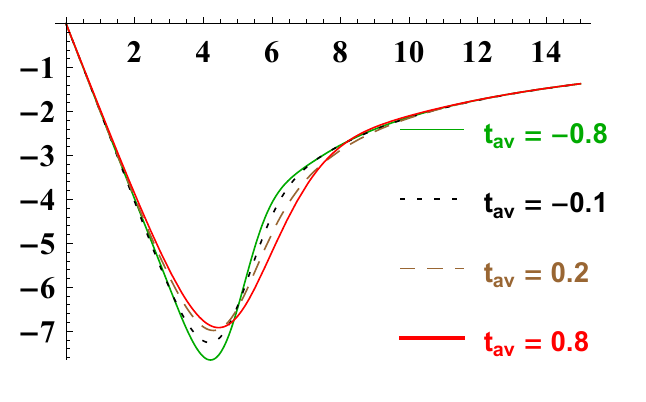}
 \caption{ $\tilde{\rho}$ as a function of $\omega$ for $\vert \mathbf{k} \vert=4.5$ and $\alpha=0.2$. Here, $t_i \sim -0.66$, $t_f \sim 0.66$, $t_{\rm av(i)} \sim -0.8$, $t_{ \rm av (f)} \sim 0.8$}
\label{fig:c4}
\end{minipage}
\begin{minipage}{0.1\textwidth}

\end{minipage}
\begin{minipage}{.45\textwidth}

\hspace{0.8cm} \includegraphics[scale=0.95]{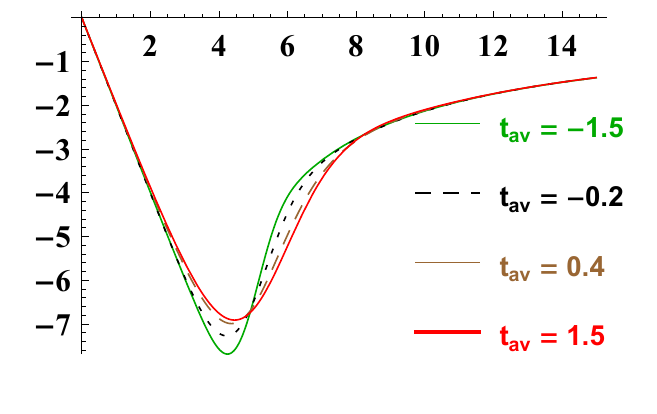}
   \caption{$\tilde{\rho}$ as a function of $\omega$ for $\vert \mathbf{k} \vert=4.5$ and $\alpha=0.5$. Here, $t_i \sim -1.55$, $t_f \sim 1.55$, $t_{\rm av(i)} \sim -1.5$, $t_{\rm av (f)} \sim 1.5$}
  \label{fig:d4}
\end{minipage}
\end{figure}

\begin{figure}[t]
\centering
\includegraphics[scale=1]{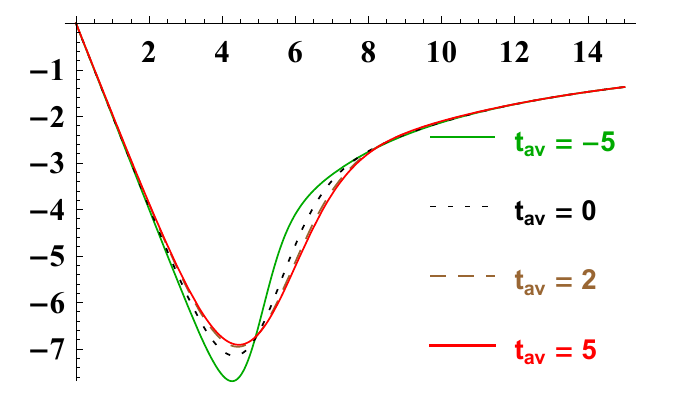}
\caption{$\tilde{\rho}$ as a function of $\omega$ for  $\vert \mathbf{k} \vert=4.5$ and $\alpha=2$. Here, $t_i \sim -6.6$, $t_f \sim 6.6$, $t_{\rm av(i)} \sim -5.0$, $t_{\rm av (f)} \sim 5.0$ }
\label{fig:e4}
\end{figure}

We explore the regime $\vert \mathbf{k} \vert > M_{\rm in}$ by choosing $\vert \mathbf{k} \vert =4.5$. The plots are presented in Figs. \ref{fig:a4}, \ref{fig:c4}, \ref{fig:d4} and \ref{fig:e4}. The patterns of thermalisation are summarised in Table \ref{k=4.5}. As observed in the previous case, adding momentum suppresses kink formation in the spectral function. We now also find that it suppresses late time persistent oscillations at $\alpha = 0.05$. The pattern is mildly non-adiabatic for $\alpha = 0.05$ (see Fig.~\ref{fig:a4}) and adiabatic for higher values (see Figs. \ref{fig:c4}, \ref{fig:d4} and \ref{fig:e4}). Advanced time-dependence still appears at $\alpha = 0.05$. 

Once again, by adiabatic we actually mean distortion-free, i.e. the spectral functions at intermediate times take almost thermal forms at instantaneous effective temperatures. Since, we cannot explore lower values of $\alpha$ with sufficient numerical accuracy, we cannot assert surely that the persistent oscillations never occur even in the instantaneous quench limit. Nevertheless, we can expect that at higher values of  $\vert \mathbf{k} \vert$ we only have mild non-adiabatic behaviour for very small values of $\alpha$ and the adiabatic pattern for higher values of $\alpha$. 


It is to be noted that in the case of $|\mathbf{k}|=4.5$ the height of the spectral peak always decreases. This is consistent with the behaviour of the thermal spectral functions at the instantaneous effective temperatures.

\subsection{Top-down or bottom-up?}
It is interesting to ask the question whether the patterns of thermalisation can be classified as top-down or bottom-up. At a fixed value of $\vert \mathbf{k} \vert$ and $\alpha$, this amounts to asking whether higher or lower frequency modes thermalise faster. In the previous subsection, our \textit{operational} definition of the time of thermalisation $t^\alpha_{\rm av(f)}$ has been such that the maximal deviation of $ \rho (\omega, t_{\rm av}, \mathbf{k})$ from its final equilibrium value $\rho_{\rm f}(\omega, \mathbf{k})$ should be less than 1 percent for $t_{\rm av} > t^\alpha_{\rm av(f)}$. This definition is blind to the $\omega-$dependence of $ \rho (\omega, t_{\rm av}, \mathbf{k})$. In order to understand $\omega$ dependence of thermalisation time, we take a closer look at $ \rho (\omega, t_{\rm av}, \mathbf{k})$ at late times, i.e. when $t_{\rm av} \approx t_{\rm av(f)}^\alpha$.

\begin{figure}[t]
\centering
\subfigure[$|\mathbf{k}|=0$]{\includegraphics[width=7cm]{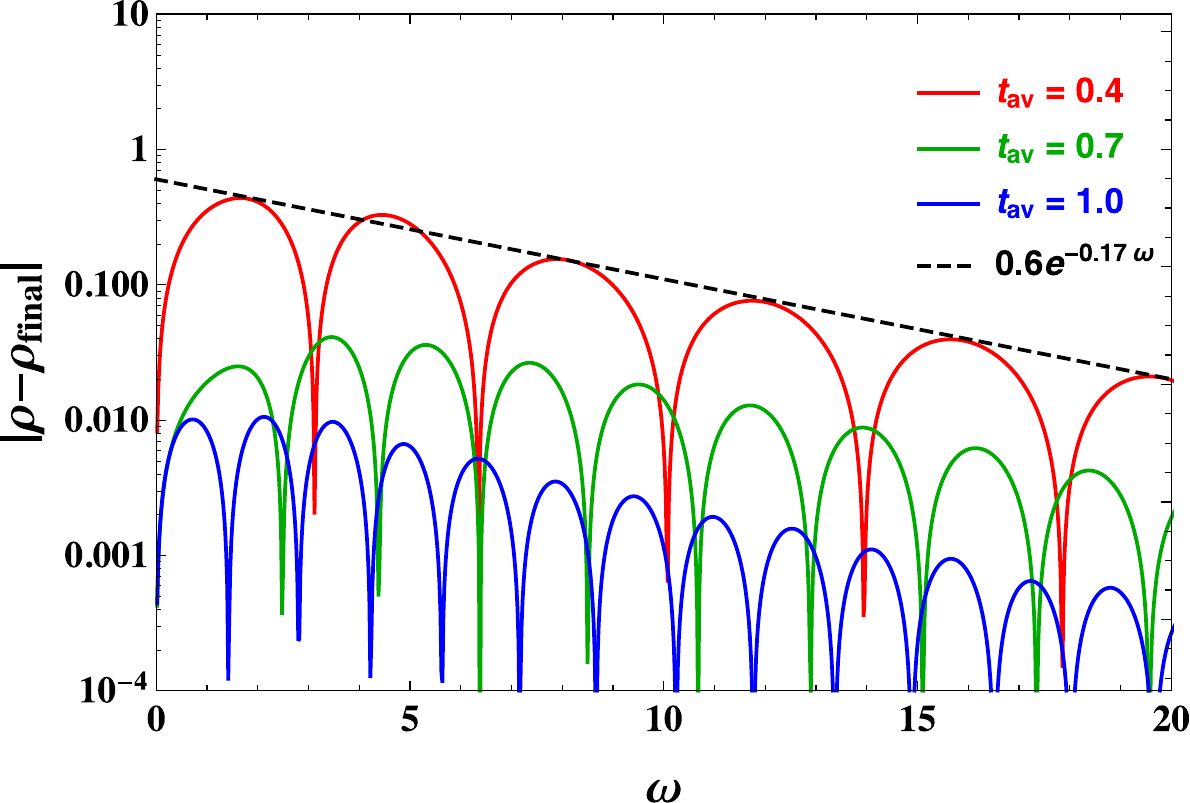}\label{fig:rhow_k0}}
\subfigure[$|\mathbf{k}|=1$]{\includegraphics[width=7cm]{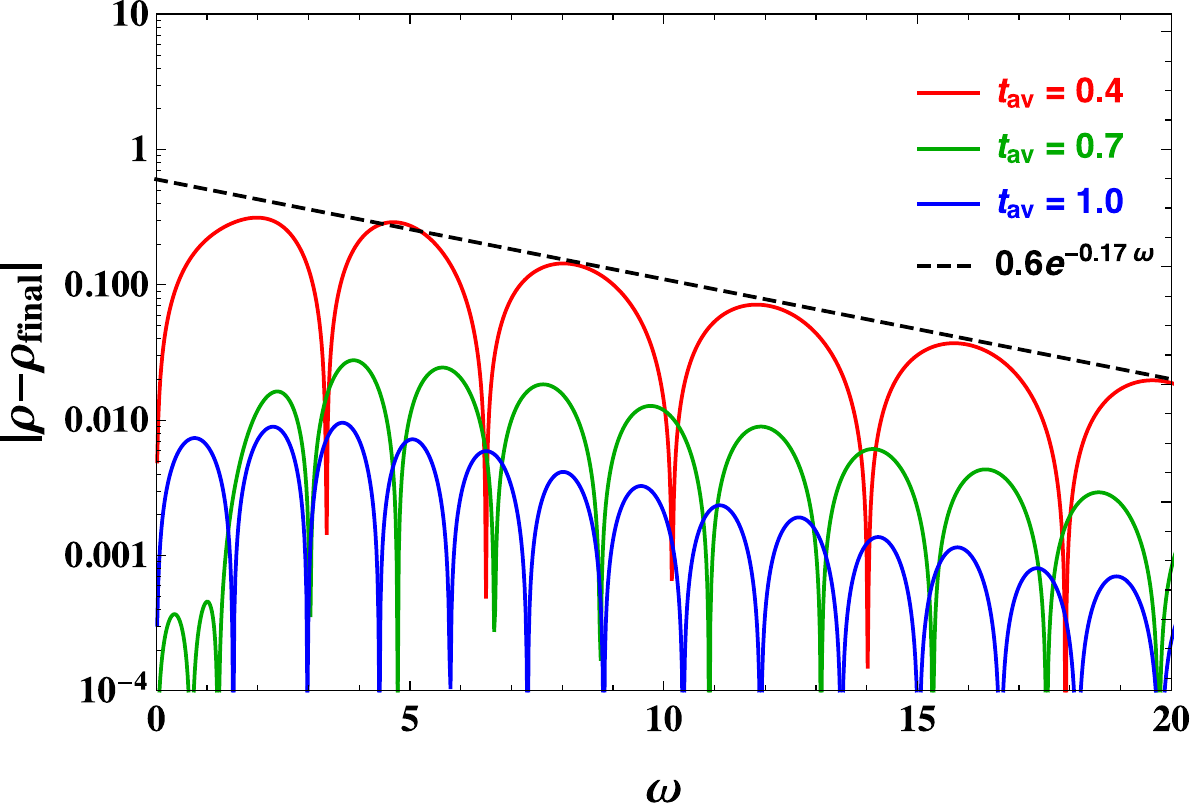}\label{fig:rhow_k1}}
\subfigure[$|\mathbf{k}|=4.5$]{\includegraphics[width=7cm]{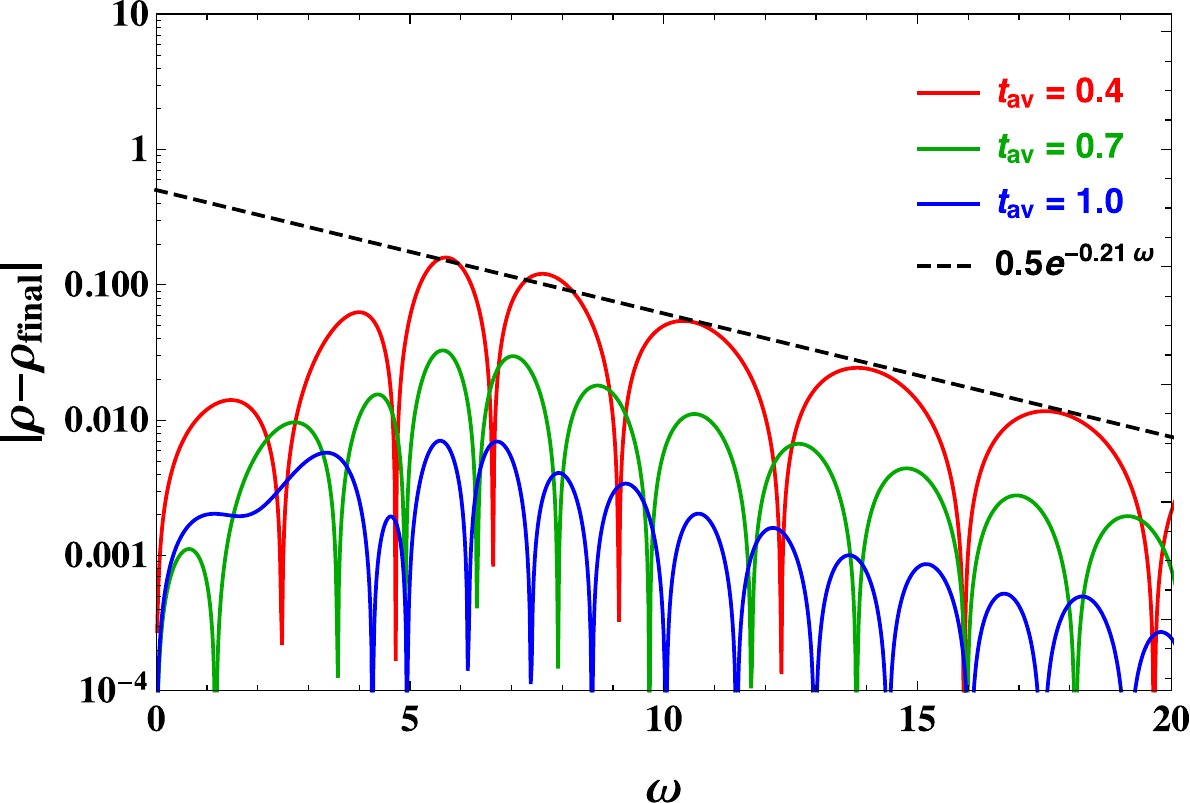}\label{fig:rhow_k45}}
\caption{Frequency dependence of $\vert\rho(\omega,t_\mathrm{av},\mathbf{k}) - \rho_{\mathrm{f}}(\omega,\mathbf{k})\vert$ for $\alpha=0.05$.}
\label{fig:rhow}
\end{figure}

\begin{figure}[t]
\centering
\includegraphics[scale=1.0]{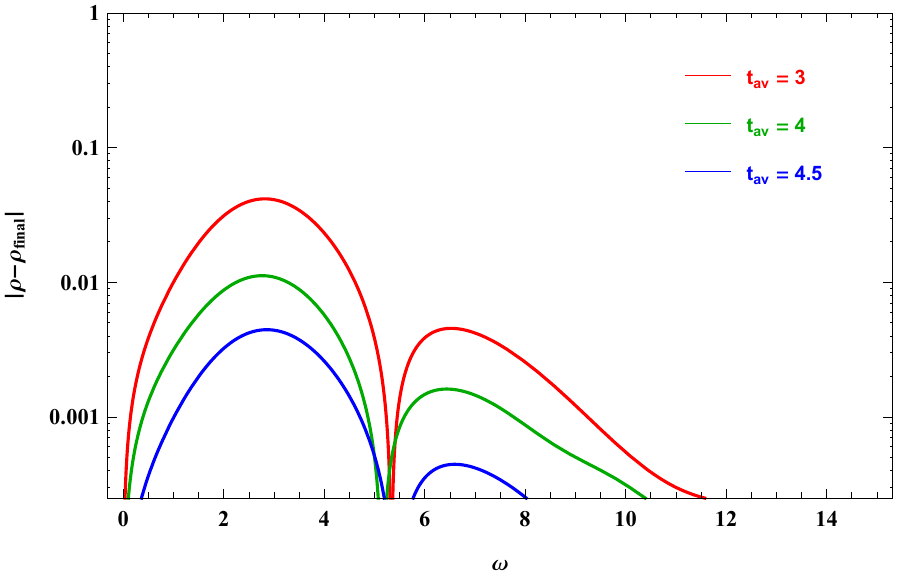}
\caption{$\vert \rho (\omega, t_{\rm av}, \mathbf{k}) - \rho_{\rm f}(\omega, \mathbf{k})\vert$ plotted in log scale as a function of $\omega$ for various values of $t_{\rm av}$ at $\alpha = 2$ and $\vert \mathbf{k} \vert = 0$ }
\label{fig:logplot2}
\end{figure}

We investigate plots of $\vert \rho (\omega, t_{\rm av}, \mathbf{k}) - \rho_{\rm f}(\omega, \mathbf{k})\vert$ as a function of $\omega$ in log scale for various values of $t_{\rm av}$ in the instantaneous quench limit $\alpha = 0.05$ for $\vert\mathbf{k}\vert=0$ (see Fig. \ref{fig:rhow_k0}),  $\vert\mathbf{k}\vert=1$ (see Fig. \ref{fig:rhow_k1}) and $\vert\mathbf{k}\vert=4.5$ (see Fig. \ref{fig:rhow_k45}). In case of $\alpha = 2$ (see Fig. \ref{fig:logplot2}), we have plotted the same quantity for $\vert\mathbf{k}\vert=0$. 

For $\alpha=0.05$, the pattern is globally top-down, but a closer look reveals complexities. There is uneven distribution of spectral weight for frequencies $0<\omega<10$.
This is readily visible from the plots for $t_{\rm av} = 0.7, 1.0$ in Figs. \ref{fig:rhow_k0}, \ref{fig:rhow_k1} and \ref{fig:rhow_k45}. However, the overall spectral weight seems to be decreasing slowly (with a very mild slope in the log plot which is $-0.17$ for $\vert \mathbf{k} \vert = 0$ and $1$, and $-0.21$ for $\vert \mathbf{k} \vert = 4.5$) at higher frequencies $\omega>10$. Such log-plots of the Wigner transformed spectral function are not readily available in the literature.\footnote{Note that if we plot $G_R(t, t_0)$ for a fixed $t_0$, we simply reproduce the one-point function $f_1(t)$ essentially. Therefore, this should follow the quasinormal mode behaviour as clearly demonstrated in Fig. \ref{fig:GR_k0a005}. However, after the Wigner transformation, the behaviour is not expected to be the same.} For extremely high frequencies, the overall spectral weight decays surely because in this region the spectral function always remains time-independent and is as in the vacuum. When $\alpha = 2$, i.e. when we are in the adiabatic limit, the oscillations are suppressed and there is more rapid decay of the spectral weight at higher frequencies.  In this case, we may proclaim the behaviour to be more of the top-down type, although strictly speaking it depends on which window of frequencies we are looking at. 

\begin{figure}[t]
\centering
\subfigure[$|\mathbf{k}|=0$]{\includegraphics[width=7cm]{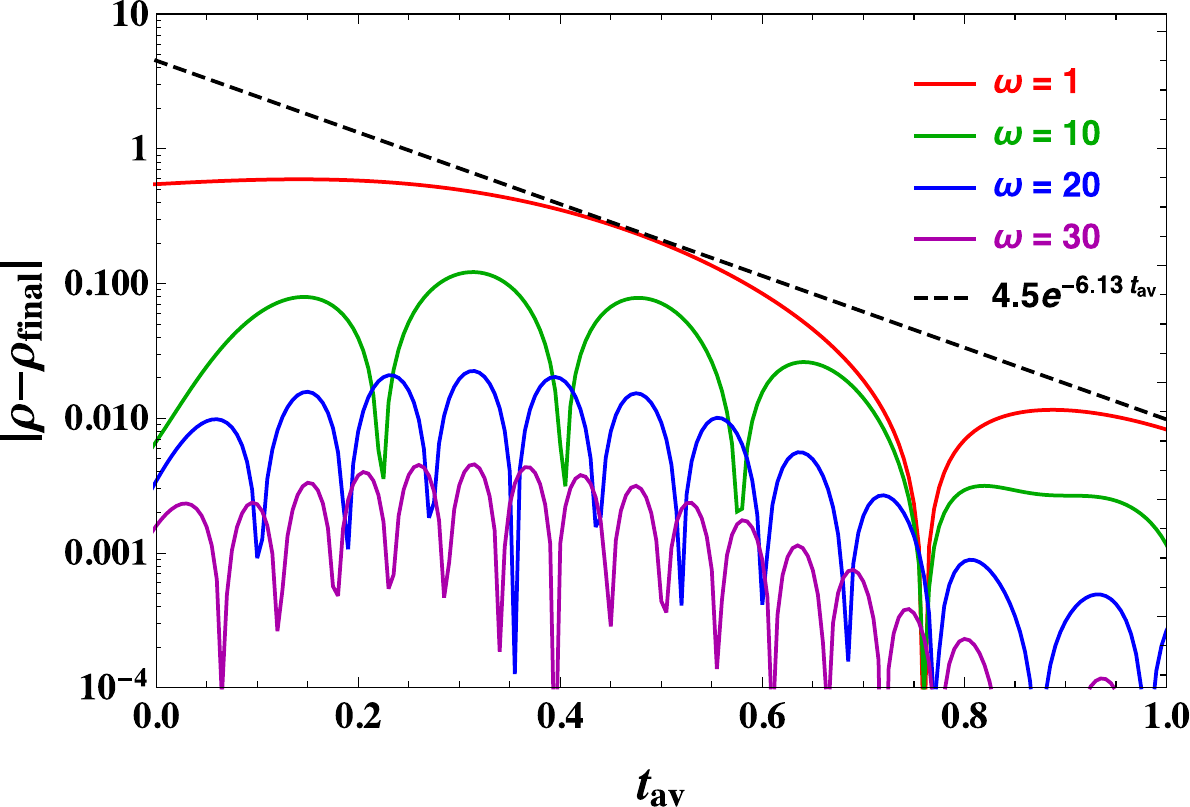}\label{fig:rhot_k0}}
\subfigure[$|\mathbf{k}|=1$]{\includegraphics[width=7cm]{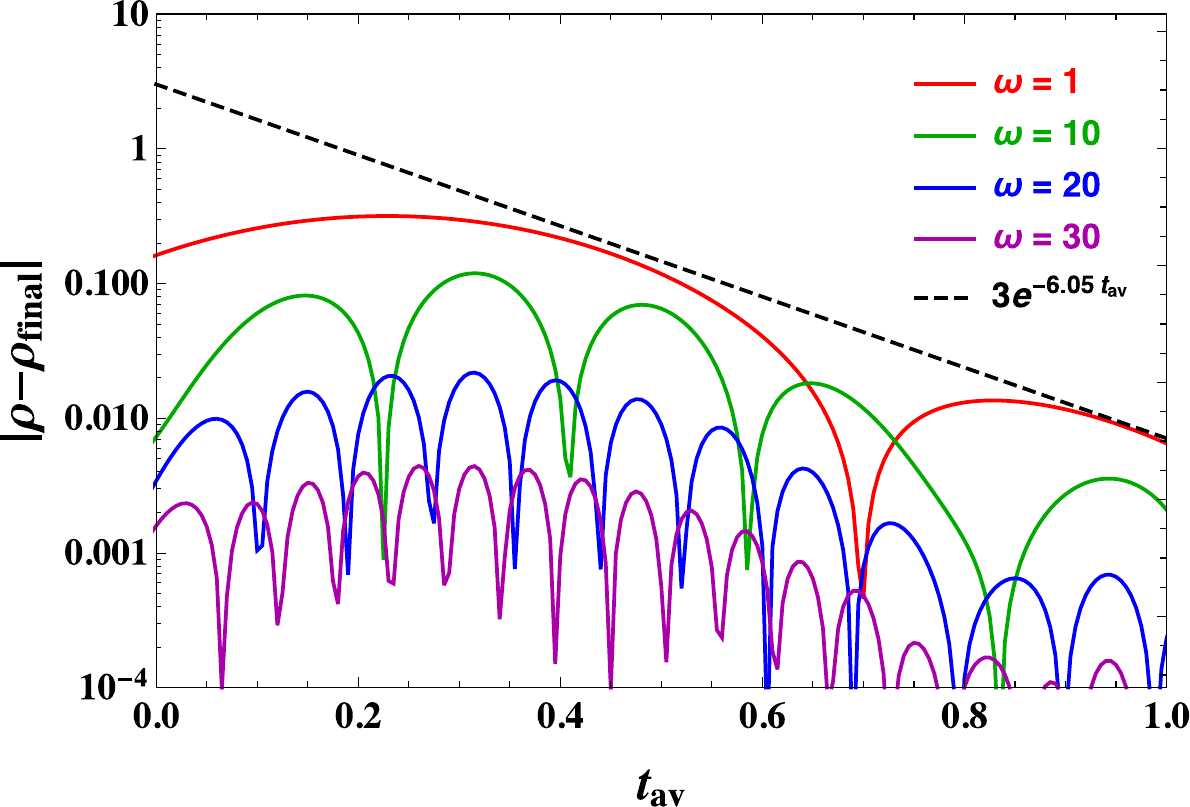}\label{fig:rhot_k1}}
\subfigure[$|\mathbf{k}|=4.5$]{\includegraphics[width=7cm]{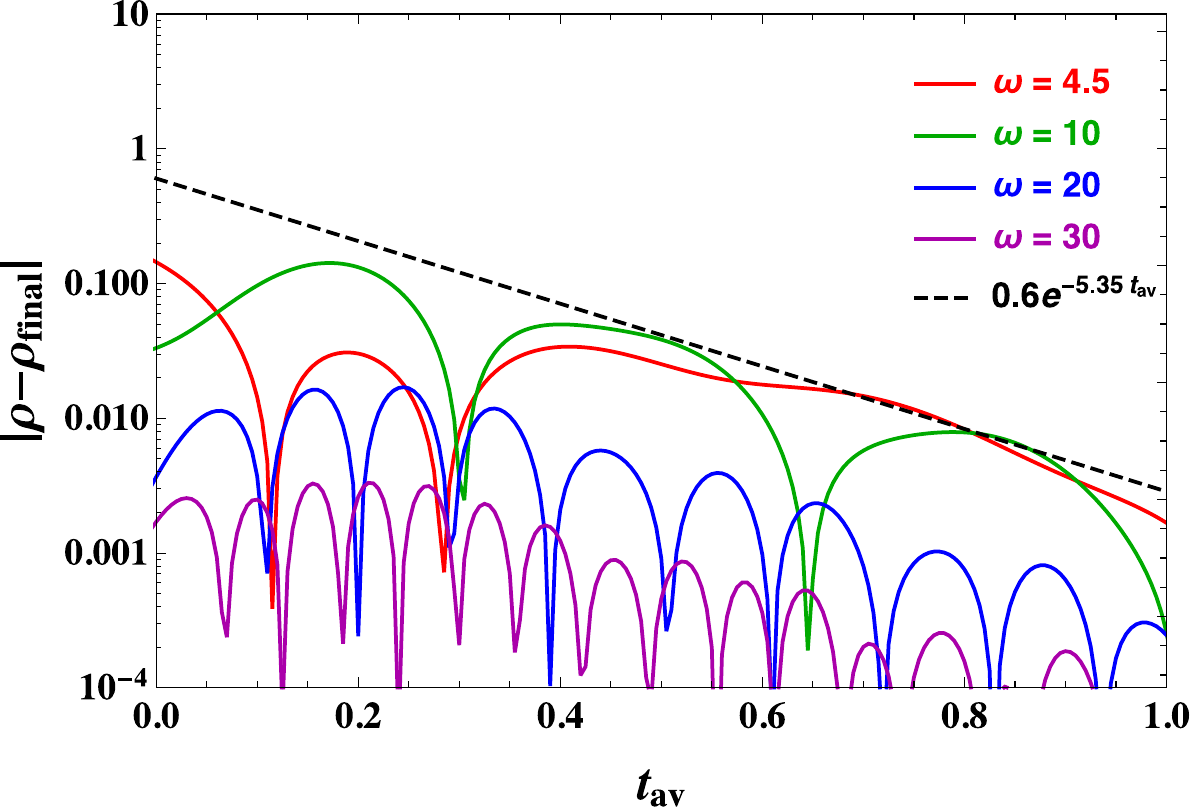}\label{fig:rhot_k45}}
\caption{Time dependence of $|\rho(\omega,t_\mathrm{av},\mathbf{k}) - \rho_{(\mathrm{f})}(\omega,\mathbf{k})|$ for $\alpha=0.05$. The black dashed line is $e^{-2 \Gamma_{(\mathrm{f})} t_\mathrm{av}}$ with $\Gamma_{(\mathrm{f})}$ being the imaginary part of the lowest quasinormal mode frequency of the final black brane. This black line should not be thought of as a fit, but rather as a benchmark for comparison. We may note that both the period and amplitude of the oscillations are modulated in complex ways for each $\omega$.}
\label{fig:rhot}
\end{figure}

In Figs. \ref{fig:rhot_k0}, \ref{fig:rhot_k1} and \ref{fig:rhot_k45}, we have plotted $\vert \rho(\omega, t_{\rm av}, \mathbf{k})- \rho_{\rm f}(\omega, \mathbf{k}) \vert$ as a function of $t_{\rm av}$ for various values of $\omega$ at $\vert \mathbf{k} \vert = 0$, $1$ and $4.5$ respectively for $\alpha = 0.05$. From these plots, it is directly visible that the patterns of thermalisation vary with $\omega$ in a complex way, and do not coincide even at late time as we would have naively expected. It is also clear from these figures that even in the high frequency region we do not get a quasinormal mode type ring-down as present in the one-point function. The reader may readily contrast these plots with Fig. \ref{fig:f1plot3} where the quasinormal mode ring-down of the one-point function has been presented. Note in the latter case, the envelope of decay in time has a linear slope in the log plot and there are also oscillations with uniform periods -- both the decay slope and oscillation period are controlled by the lowest quasinormal mode of the final black brane. In case of $\omega = 10$, $20$ and $30$, both the period and amplitude of the oscillations in the spectral function get modulated in complex ways even at late time $t_{\rm av} \approx t^\alpha_{\rm av(f)}$ as visible from the plots.

Although we find complicated patterns of thermalisation of the spectral function at various values of $\omega$, it may still be possible that large scale features of time-dependence are controlled by the lowest quasinormal mode of the final black brane for all $\omega$. We need more numerical accuracy, data and computational time in order to see such features, if they exist.

\begin{figure}[t]
\centering
\includegraphics[scale=1.0]{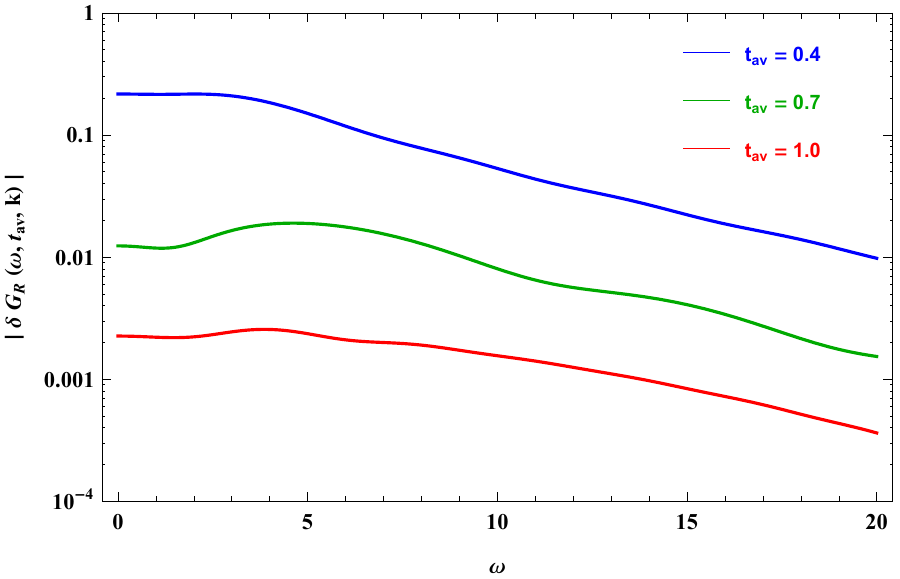}
\caption{$\vert \delta G_R (\omega, t_{\rm av}, \mathbf{k} = 0)\vert$ plotted in log scale as a function of $\omega$ for various values of $t_{\rm av}$ at $\alpha = 0.05$}
\label{fig:deltaGR}
\end{figure}

\begin{sloppypar}
Interestingly, if we plot the modulus of the full subtracted retarded propagator $\vert \delta G_R (\omega, t_{\rm av}, \mathbf{k})\vert$ on a log scale as before, where $\delta G_R  = G_R - G_{R\,\rm f}$ with  $ G_{R\,\rm f}$ being the final thermal form of the retarded propagator, then even at $\alpha = 0.05$ and $\vert \mathbf{k}\vert= 0$ the oscillations are suppressed (see Fig. \ref{fig:deltaGR}). Furthermore, at late time the contribution remains more or less uniform up to frequencies $\omega \approx M_{\rm f}$ and then seems to decrease \textit{exponentially}. Thus how close the pattern is towards a top-down description may depend on our actual observable as well as on the specific window of frequencies at which we are making our observations. 
\end{sloppypar}

In order to understand better how different frequency bands thermalise, the Gabor transform as utilised in \cite{Chesler:2012zk} could be a valuable tool.\footnote{We thank the referee for pointing this out.} Although the conventional Wigner transform sheds more light on the global aspects of thermalisation, and spectral weight distribution across all frequencies as a function of time, etc. and gives direct connections with experimental results (as discussed below), other representations like Gabor transform can shed light on the time-dependent dynamics of individual frequency bands. We would like to explore such questions in the future.

\section{Conclusions}

Let us sum up lessons learnt from the results presented so far and point out some interesting questions that merit further investigation. The most important lesson is that we see clearly distinct patterns for the thermalisation of the non-equilibrium spectral function as we vary parameters determining the dynamics of the non-equilibrium state even in cases where the one-point functions thermalise in a universal way via a thermal relaxation mode. Indeed, it has been observed before that even if we go beyond linear response regime in holographic theories, the non-equilibrium evolution of the one-point function of a scalar operator is very well approximated by thermal linear response to a very good degree of approximation even close to initial stage when the system is far from equilibrium \cite{Bhaseen:2012gg,Heller:2013oxa}. Furthermore, in \cite{Das:2014hqa} it has been shown that even at \textit{early} times, i.e. before the endpoint of the quench, the one-point functions and space-like separated correlations behave in certain universal ways irrespective of the quench protocol  in general conformal field theories. In our example constituting of states approximated by AdS-Vaidya spacetimes, the one point function do not show visibly distinct pattern as it makes transition from the thermal response of the initial black hole at early time to that of the final black hole at late time. However, the non-equilibrium spectral function \textit{knows} about what is going on in the bulk more clearly. Depending on the compactness of the shell driving the expansion of the non-equilibrium horizon (reflecting the duration of the quench), it does show very different patterns of thermalisation.

The second lesson learnt is that in order to find different patterns of thermalisation of two-point functions, we should go beyond geodesic approximation in the bulk, and of course also vary the parameters determining the non-equilibrium dynamics in the state. Going beyond the geodesic approximation, we find kink formation and advanced time-dependence in particular (as discussed before) and also finer structure in the late-time oscillations when they exist. Our method is certainly well suited to investigate beyond the simple example considered here. It can be readily applied even to situations where the holographic geometry dual to the nonequilibrium state can be found only numerically. It is worth noting that some of the most generic features of non-equilibrium dynamics including quasinormal modes in the \textit{background} gravitational geometry at late times are not present in the simple examples we have investigated here. 

The most experimentally relevant set-up that can be modeled via holography is related to solid state pump-probe spectroscopy (see for instance \cite{Segre:2001aa} for results of such experiments on cuprates). In this case, the material is driven away from equilibrium by a strong laser pulse, and a weak laser pulse is sent after an adjustable delay (which can be as short as an attosecond). The latter gets modulated by the medium revealing information about the time-dependence of the current-current retarded propagator. Holographically, the non-equilibrium state can be readily modelled by a dual geometry that is excited by a specific source at the boundary (which needs to be obtained numerically). Our prescription can then be applied to obtain the non-equilibrium retarded propagator. We would like to investigate such holographic set-ups in the future.

Recently in \cite{Stefan}, it has been shown that even in the geodesic-like approximation the time-dependent behaviour of non-local observables like entanglement entropy depend strongly on the initial conditions given by the geometry of colliding shockwaves in holographic models of heavy-ion collisions. Such qualitative features are not prominently visible in the one-point functions, as for instance in the energy-momentum tensor. It should be interesting to understand if the time-dependence of the holographic spectral function can indeed reveal the detailed geometry of the colliding shock waves. In that case, it would also be interesting to understand how these time-dependent spectral functions can be measured in heavy-ion collision experiments.

A bigger question could be as follows: can we make a broad classification of patterns of thermalisation of the non-equilibrium Schwinger-Keldysh propagators in strongly interacting many-body systems? This is analogous to classifying hydrodynamic flows qualitatively based on the Reynolds number. The analogue of the Reynolds number in this case should be appropriate combinations of external parameters driving the nonequilibrium dynamics and intrinsic parameters (like masses and couplings of the bulk theory, or equivalently the spectrum of primary operators and their three-point functions in the dual CFT) specifying the system. Indeed holography can lead us to such a broad classification which may be compared with experimental results.

Finally, we have proposed a method only for obtaining causal holographic Schwinger-Keldysh propagators at and away from equilibrium. We need to extend our method further for obtaining the other Schwinger-Keldysh propagators (specifically the Feynman propagator). This will give us insight into generalisation of fluctuation-dissipation relations away from equilibrium, and complete information about thermalisation and decoherence. In order to achieve this, we need to go beyond purely causal response by making the boundary sources in holography dynamical in a self-consistent way as in semi-holography (see \cite{Iancu:2014ava,Mukhopadhyay:2015smb} in particular). We will like to generalise our method using the latter framework in future.

\section*{Acknowledgments}
We thank Hans Bantilan, Justin David, Christian Ecker, Sebastian Fischetti, Matthias Kaminski, Johannes Oberreuter, Marios Petropoulos, Florian Preis, Anton Rebhan, Paul Romatschke, Konstantinos Siampos, Stefan Stricker, Christopher Rosen, Toby Wiseman and Jackson Wu for fruitful discussions and comments. We also thank Christian Ecker and Stefan Stricker for comments on the manuscript. The work of T.I. was supported by the Department of Energy, DOE award No. DE-SC0008132. A.\ Mukhopadhyay acknowledges support from a Lise Meitner fellowship
of the Austrian Science Fund (FWF), project no.\ M1893-N27.

 \appendix
\section{The case of $ \vert \mathbf{k} \vert = 0.5$}
\label{appendixA}
\begin{figure}[h!]
\centering
\includegraphics[scale=0.95]{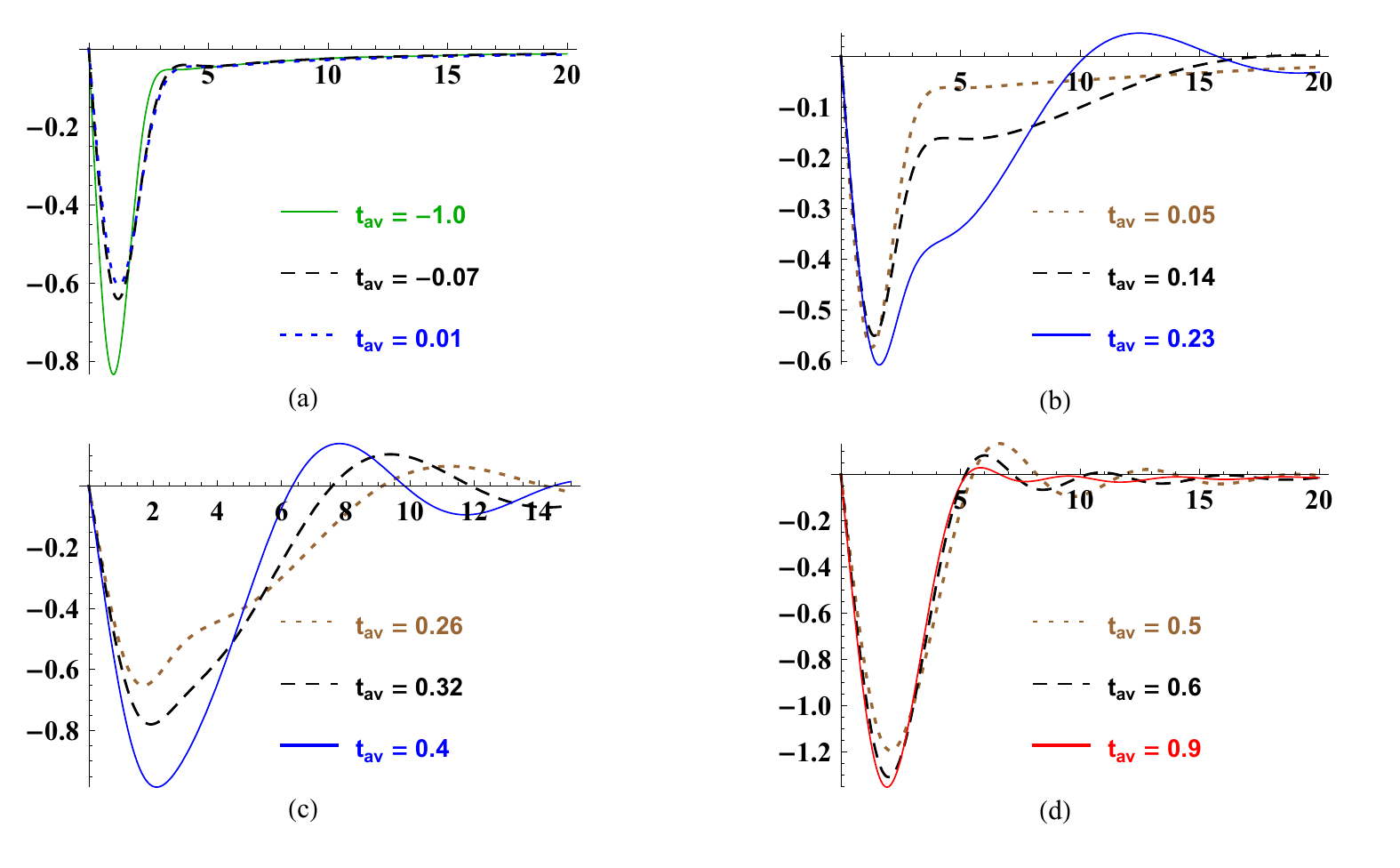}
 \caption{$\tilde{\rho}$ as a function of $\omega$ for $|\mathbf{k}|=0.5$ and $\alpha=0.05$. Here, $t_i \sim -0.17$, $t_f \sim 0.17$, $t_{\rm av(i)} \sim -1.0$, $t_{\rm av (f)} \sim 0.9$}
\label{fig:a2}
\end{figure}

\begin{figure}[t]
\centering
\includegraphics[scale=0.95]{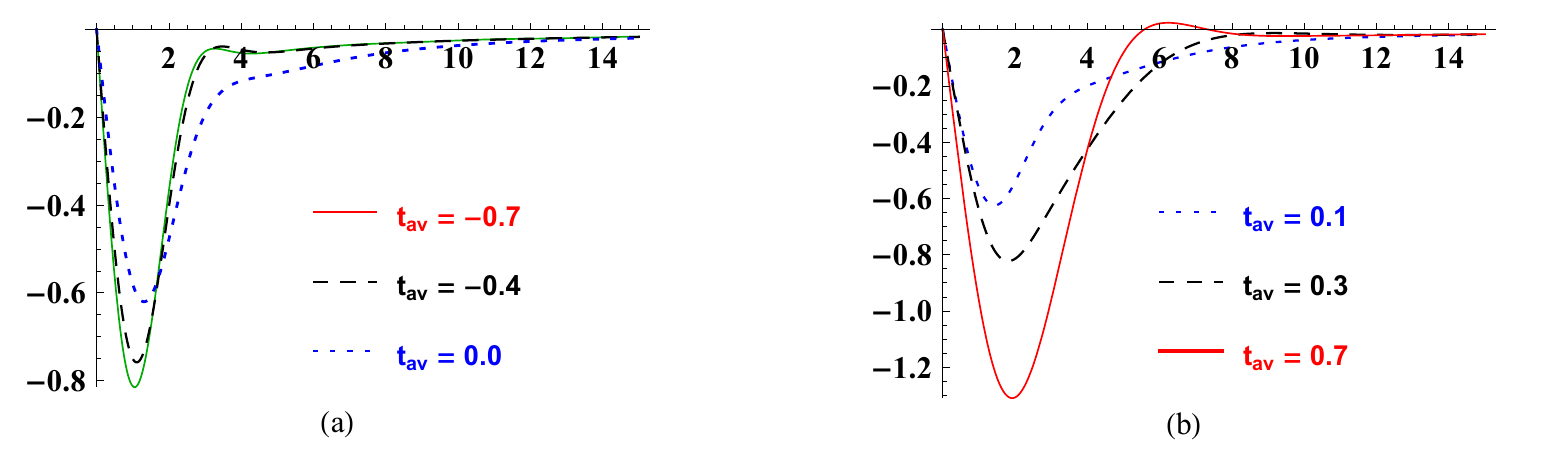}
\caption{$\tilde{\rho}$ as a function of $\omega$ for $|\mathbf{k}|=0.5$ and $\alpha=0.2$. Here, $t_i \sim -0.66$, $t_f \sim 0.66$, $t_{\rm av(i)} \sim -0.75$, $t_{\rm av (f)} \sim 0.75$}
\label{fig:c2}
\end{figure}

\begin{figure}[t]
\centering
\includegraphics[scale=0.95]{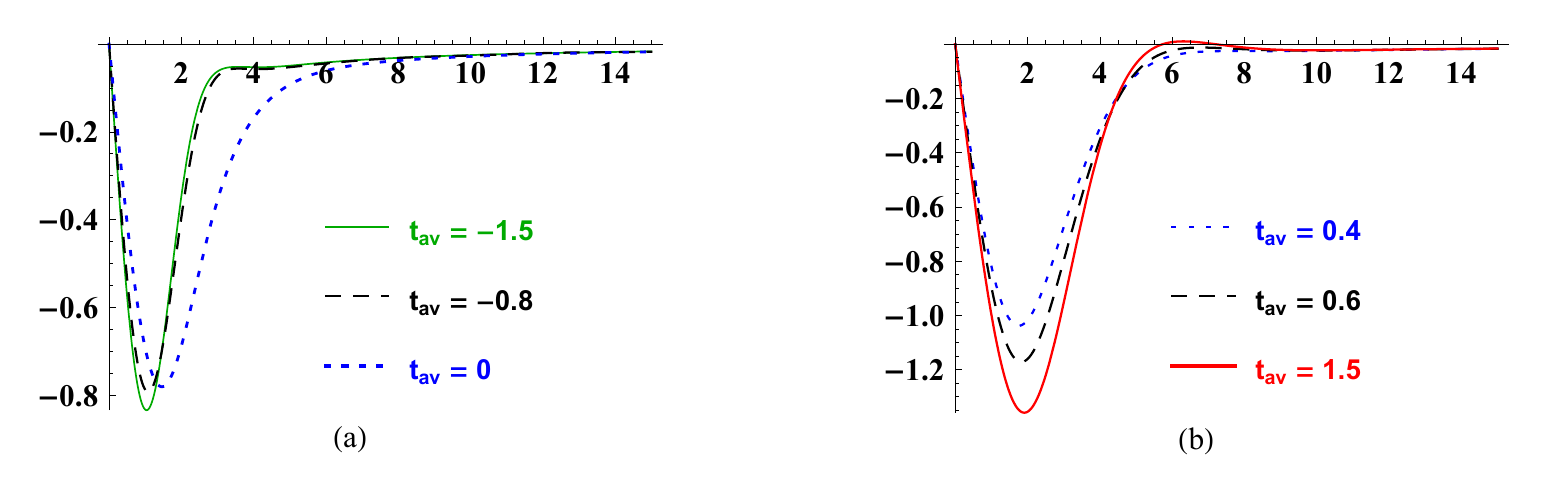}
\caption{$\tilde{\rho}$ as a function of $\omega$ for $|\mathbf{k}|=0.5$ and $\alpha=0.5$. Here, $t_i \sim -1.55$, $t_f \sim 1.55$, $t_{\rm av(i)} \sim -1.5$, $t_{\rm av (f)} \sim 1.5$ }
\label{fig:d2}
\end{figure}

\begin{figure}[t]
\centering
\includegraphics[scale=1]{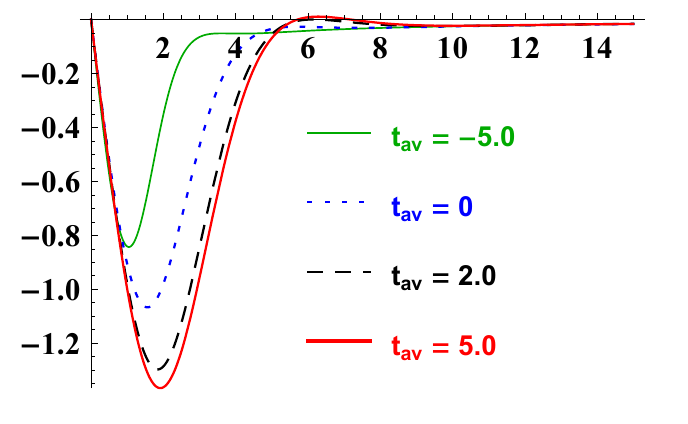}
\caption{$\tilde{\rho}$ vs. $\omega$ as a function of $|\mathbf{k}|=0.5$ and $\alpha=2.0$. Here, $t_i \sim -6.6$, $t_f \sim 6.6$, $t_{\rm av(i)} \sim -5.0$, $t_{\rm av (f)} \sim 5.0$}
\label{fig:e2}
\end{figure}

We have observed in Section \ref{Results} that the patterns of thermalisation of the spectral function for $\vert \mathbf{k} \vert=0.5$ are similar to those in the zero momentum case. Here we present the plots for the same values of $\alpha$. Figs. \ref{fig:a2}, \ref{fig:c2},  \ref{fig:d2} and Fig. \ref{fig:e2} indeed show identical features of thermalisation as given in Table \ref{k=0} for $\alpha = 0.05$, $0.2$, $0.5$ and $2.0$ respectively. 
\vskip -0.9cm



\clearpage 

\bibliographystyle{JHEP} 

\bibliography{NoneqSpecFunc}

\end{document}